\RequirePackage{fix-cm}
\documentclass[smallextended]{svjour3}       
\smartqed  
\usepackage[utf8]{inputenc} 
\usepackage{amsmath,amssymb,hyperref,graphicx,comment}

\usepackage{tikz}
\usetikzlibrary{positioning}
\pgfdeclarelayer{bg}    
\pgfsetlayers{bg,main}  

\newcommand{\rr}{\mathbf{r}}
\newcommand{\pp}{\mathbf{p}}
\newcommand{\diff}{\,\mathrm{d}}

\newcommand{\uu}{\mathbf{u}}

\newcommand{\dr}{\,\diff\rr}

\newcommand{\LL}{\mathbf{L}}

\newcommand{\upi}{{^\uparrow \pi}}
\newcommand{\dpi}{{^\downarrow \pi}}
\newcommand{\ppi}{\dpi}
\newcommand{\pim}[2]{\dpi^{#1\,}_{\,#2}}

\newcommand{\xx}{\mathbf{x}}
\newcommand{\yy}{\mathbf{y}}

\newcommand{\XX}{\mathbf{X}}
\newcommand{\YY}{\mathbf{Y}}

\newcommand{\uS}{{^\uparrow S}}
\newcommand{\dS}{{^\downarrow S}}
\newcommand{\uE}{{^\uparrow E}}
\newcommand{\dE}{{^\downarrow E}}
\newcommand{\uL}{{^\uparrow L}}
\newcommand{\uLL}{{^\uparrow \mathbf{L}}}
\newcommand{\dLL}{{^\downarrow \mathbf{L}}}
\newcommand{\dL}{{^\downarrow L}}
\newcommand{\dLr}{{^\downarrow \tilde{L}}}
\newcommand{\dLLr}{{^\downarrow \tilde{\LL}}}
\newcommand{\uPB}[2]{{^\uparrow\left\{ #1, #2 \right\}}}
\newcommand{\dPB}[2]{{^\downarrow\left\{ #1, #2 \right\}}}
\newcommand{\dPBc}[2]{{^\downarrow\left\{ #1, #2 \right\}}^*}
\newcommand{\Lagr}{\mathcal{L}}

\newcommand{\Hami}{\mathcal{H}}
\newcommand{\dyy}{\dot{\yy}}
\newcommand{\dy}{\dot{y}}
\newcommand{\zz}{\mathbf{z}}

\newcommand{\uXi}{{^\uparrow \Xi}}
\newcommand{\dXi}{{^\downarrow \Xi}}

\newcommand{\uM}{{^\uparrow M}}
\newcommand{\dM}{{^\downarrow M}}
\newcommand{\dMM}{{^\downarrow \MM}}

\newcommand{\uEb}{{^\uparrow \bar{E}}}

\newcommand{\qq}{\mathbf{q}}
\newcommand{\nx}{\hat{x}}
\newcommand{\nxx}{\hat{\xx}}
\newcommand{\nS}{\hat{S}}
\newcommand{\nXi}{\hat{\Xi}}
\newcommand{\nM}{\hat{M}}
\newcommand{\MM}{\mathbf{M}}
\newcommand{\nMM}{\hat{\MM}}
\newcommand{\nq}{\hat{q}}
\newcommand{\np}{\hat{p}}

\newcommand{\ppsi}{\boldsymbol{\psi}}
\newcommand{\PPhi}{\boldsymbol{\Phi}}
\newcommand{\BB}{\mathbf{B}}

\newcommand{\pd}[2][]{\frac{\partial #1}{\partial #2}}
\newcommand{\pdv}[2][]{\frac{\delta #1}{\delta #2}}

\newcommand{\VF}{\mathfrak{X}}
\newcommand{\MMM}{\mathcal{M}}
\newcommand{\NNN}{\mathcal{N}}

 \usepackage{mathptmx}      

\begin{document}

\title{Generalization of the dynamical lack-of-fit reduction
}
\subtitle{From GENERIC to GENERIC}


\author{Michal Pavelka \and Václav Klika \and Miroslav Grmela}


\institute{M. Pavelka\at
           Mathematical Institute, Faculty of Mathematics and Physics, Charles University, Sokolovsk\'{a} 83, 186 75 Prague, Czech Republic \\
              \email{pavelka@karlin.mff.cuni.cz}           
           \and
           V. Klika\at
Czech Technical University in Prague, Department of Mathematics – FNSPE, Trojanova 13, 120 00 Prague, Czech Republic
           \and
	   M. Grmela\at
\'{E}cole Polytechnique de Montr\'{e}al, C.P.6079 suc. Centre-ville, Montr\'{e}al, H3C 3A7,  Qu\'{e}bec, Canada
}

\date{Received: date / Accepted: date}

\maketitle

\begin{abstract}
	The lack-of-fit statistical reduction, developed and formulated first by Bruce Turkington,  is a general method taking Liouville equation for probability density (detailed level) and transforming it to reduced
dynamics of projected quantities (less detailed level). In this paper the method is generalized. The Hamiltonian Liouville equation is replaced by an arbitrary Hamiltonian evolution combined with gradient dynamics (GENERIC), the Boltzmann
entropy is replaced by an arbitrary entropy, and the kinetic energy by an  arbitrary energy. The gradient part is a generalized gradient dynamics generated by a dissipation potential. The reduced
evolution of the projected state variables is shown to preserve  the GENERIC structure of the original (detailed level) evolution. The dissipation potential is obtained by solving a Hamilton-Jacobi equation. In summary, the lack-of-fit reduction can start with GENERIC and obtain GENERIC for the reduced state variables.

\end{abstract}

\newpage

\tableofcontents

\paragraph
\textit{"With thermodynamics, one can calculate almost everything crudely; with kinetic theory, one can calculate fewer things, but more accurately; and with statistical mechanics, one can calculate almost nothing exactly."}

\hspace{10pt}\textit{    — Eugene Wigner}

\section{Introduction}

Imagine a very complex detailed dynamics of state variables $\xx$ on manifold (or vector space) $\MMM$ given by vector field $\XX\in\VF(\MMM)$, i.e. with evolution equations
\begin{equation}\label{eq.dx}
	\dot{\xx} = \XX|_\xx.
\end{equation}
Assume now that this detailed (upper-level) dynamics is too complex to be solved while keeping all the details, and that experimental observations indicate  the existence of  an autonomous  lower-dimensional dynamics that  displays  important
features of the detailed dynamics. An investigation of this type of reductions is the primary objective of thermodynamics and statistical physics.

We shall focus on evolution equations in non-equilibrium thermodynamics like classical irreversible thermodynamics \cite{dGM}, extended irreversible thermodynamics \cite{Jou-EIT}, thermodynamics with internal variables
\cite{Van-Berezovski} and the General Equation of Non-Equilibrium Reversible-Irreversible Coupling (GENERIC) \cite{go,og,hco,PKG}. In particular, we assume that the detailed evolution possess the structure of GENERIC, i.e.
the vector field consists of a reversible Hamiltonian part (generated by a Poisson bracket and energy) and an irreversible gradient part (generated by a dissipation potential and entropy). The reason for seeking the reversible evolution in Hamiltonian form is that mechanics is the reversible part and mechanics is Hamiltonian geometry (at least in the Galilean relativistic setting, see \cite{hco} for general relativistic considerations). The irreversible part represents thermodynamics, which means approach to equilibrium driven by a potential. Gradient dynamics is of course not the only possibility, but a suitable and general one, see Sec. \ref{sec.GENERIC} for more details and context. 

Consider now a less detailed (lower) manifold with state variables $\yy\in\NNN$ given by a projection $\dpi:\MMM\rightarrow\NNN$.
This manifold expresses some overall characteristics of the detailed manifold $\MMM$. Dynamics on the lower level of description is induced by the upper dynamics,
\begin{equation}
	\dot{y}^a = \frac{\partial \dpi^a}{\partial x^i}\dot{x}^i,
      \end{equation}
which is the reason why we call $\dpi$ a projection.
These evolution equations are however typically not in a closed form (expressed only in terms of the lower-level variables). In other words, there are many points $\xx$ projected to a single $\yy$. Consequently,  there are many vectors attached to the points to be projected to a single unknown vector attached to $\yy$. How to construct the vector field on $\YY\in\VF(\NNN)$ that generates the time evolution in $\NNN$ 
\begin{equation}
	\dot{\yy} = \YY|_{\yy},
\end{equation}
displaying the important features of the time evolution in $\MMM$ generated by
$\XX\in\VF(\MMM)$?

Let us now shortly recall the methods of thermodynamic reductions following \cite{Paco-Miroslav}.
Perhaps the simplest method of projecting $\XX\in\VF$ on $\YY\in \VF(\NNN)$ is provided by the principle of Maximum Entropy (MaxEnt), \cite{jaynes,redext,PKG,netocny2002}.  Picking one point $\yy\in\NNN$, there is an associated point $\upi(\yy)\in\MMM$ that corresponds to the maximum value of entropy subject to the constraint that $\dpi(\upi(\yy))=\yy$. The entropy is the potential driving the time evolution in $\MMM$ to the equilibrium.
Taking the vector attached to the point $\upi(\yy)$ and projecting it to the tangent space of $\yy$, we obtain a vector field $\YY\in\VF(\NNN)$
\begin{equation}
	Y^a(y) = \frac{\partial \dpi^a}{\partial x^i}\Big|_{x(y)}X^i(x(y)).
\end{equation}
But this vector field has a drawback. Typically, the trajectories obtained by solving evolution equations $\dot{y} = \YY$ approximate poorly the trajectories  on the $\MMM$ manifold, see e.g. \cite{Turkington}. For instance, only reversible evolution is typically obtained on $\NNN$ when evolution on $\MMM$ is irreversible. This is because the entropy used in MaxEnt addresses the approach to the thermodynamic equilibrium and not to the reduced time evolution taking place on $\NNN$.
A more precise approximation is needed to obtain on $\NNN$  a more faithful approximation of the time evolution on $\MMM$, see e.g. \cite{gorban}, \cite{DynMaxEnt}. MaxEnt is often called a static reduction, as it well approximates the detailed manifold $\MMM$ by the less detailed manifold $\NNN$, but to have evolution on $\NNN$ one needs a dynamical reduction giving also a good approximation of the vector field on $\MMM$ by a vector field on $\NNN$.

A classical example of dynamical reduction is the Chapman-Enskog asymptotic expansion \cite{Chapman-Cowling,ADER,dGM}. Let $\MMM$ be the state space of kinetic theory (i.e., the physical system under investigation is a gas and $\xx$ is the one-particle distribution function) and $\NNN$ is the state space of the hydrodynamics (i.e., hydrodynamic fields of density, momentum density and energy density, $\yy=(\rho,\uu,e)$).
In this case, the time evolution taking place in $\NNN$ is often well described by the Navier-Stokes-Fourier system of equations, see e.g. \cite{dGM}, obtained by the Chapman-Enskog expansion. The projection $\dpi$ is the projection on the first 5 moments of the distribution function, and the detailed Boltzmann equation (vector field $\XX$) is reduced to less detailed Navier-Stokes-Fourier equations (vector field $\YY$). The upper entropy $\uS$ is the Boltzmann entropy and it generates a lower-level entropy $\dS$, expressed by the Sackur-Tetrode relation for ideal gases \cite{callen,PKG}.
The embedding $\upi$ is the MaxEnt mapping from hydrodynamic fields to the locally Maxwellian distribution functions. The locally Maxwellian distribution functions form the local equilibrium submanifold of $\MMM$, which is isomorphic to $\NNN$. When the evolution in $\MMM$ takes place close to the local equilibrium submanifold, the evolution in $\NNN$ is close to the detailed evolution in $\MMM$. The Chapman-Enskog expansion has  however also a few drawbacks. Firstly, it relies on the a priori unknown form of asymptotic expansion, secondly, it requires the presence of dissipative terms in vector field $\XX$ and thirdly, the number of state variables is typically not changed and hence one cannot a priori specify the lower manifold $\NNN$ (although in Chapman-Enskog analysis of kinetic theory the evolution equations for several moments of the distribution function somewhat naturally appear). From the geometrical point of view, the first Chapman-Enskog approximation can bee seen as a deformation of $\NNN^{(0)}$ obtained by the MaxEnt reduction (in the case of the Boltzmann equation, elements of  $\NNN^{(0)}$ are the  local Maxwellian distributions),  with the objective to arrive at a manifold $\NNN^{(1)}$ that has the following two properties: (i) $\NNN^{(1)}$ is isomorphic to $\NNN^{(0)}$, and (ii) the vectors $\XX|_{\NNN^{(1)}}$ are  "closer" to the tangent vectors of $\NNN^{(1)}$ than the vectors $\XX|_{\NNN^{(0)}}$ are to the tangent vectors of $\NNN^{(0)}$.

Another method of constructing the reduced vector field is the Ehrenfest method developed in \cite{Ehrenfests,GK-Ehrenfest,GK-Ehrenfest2} and \cite{gorban}. The method  has the following ingredients: detailed manifold $\MMM$ equipped with entropy and with a vector field (evolution equations), manifold $\NNN$ and projection $\dpi$ from $\MMM$ to $\NNN$. MaxEnt then provides the embedding of $\NNN$ into $\MMM$ as usual. The vector field on $\MMM$ does not need to have the GENERIC structure, but it is advantageous as shown in \cite{Pavelka-LD}. The vector field $\XX\in\VF(\MMM)$ is first projected to a vector field $\YY_0\in\VF(\NNN)$ by the MaxEnt projection. This vector field, however, needs to be corrected as mentioned above. Therefore, the vector field $\XX$ is lifted to the tangent bundle $T\MMM$ and subsequently projected back to $\MMM$, which results in a smoothed vector field on $\MMM$, $ER(\XX(\MMM))$, which expresses a sort of overall motion on $\MMM$, called Ehrenfest regularization in \cite{Ehrenfest-regularization}. The same is done with the vector field $\YY_0$, which results in vector field $ER(\YY_0)\in\VF(\NNN)$. Finally, vector field $ER(\XX)$ is MaxEnt-projected to $\NNN$ and compared with $ER(\YY_0)$. A correction term is then added to $\YY_0$, forming a new vector field $\YY_1\in\VF(\NNN)$, which makes $ER(\XX)$ equal to $ER(\YY_1)$ (to a given order of relaxation time parameter). Vector field $\YY_1$ then represents the evolution on $\NNN$, its components are right hand sides of evolution equations for $y\in\NNN$. This is the Ehrenfest reduction of detailed evolution on $\MMM$. An advantage of this dynamical reduction is that it provides irreversible evolution on $\NNN$ even if evolution on $\MMM$ is purely reversible. A drawback is, however, that a constant relaxation time parameter is to be supplied and its value is not predicted by the theory.

Different method of dynamic reduction is the Dynamic MaxEnt developed in \cite{Grmela2013-CMAT,redext,DynMaxEnt}. The main idea is to first promote the conjugate variables $\xx^*$ in the GENERIC framework to independent variables, which is natural from the point of view of contact geometry \cite{ent16,PKG}. The goal is to reduce a GENERIC model for state variables on manifold $\MMM$ so that a fast variables relaxes and becomes enslaved by the remaining slower variables, $\NNN$ being the manifold of slow variables. The fast variable is first evaluated at the MaxEnt value determined by the remaining state variables. But since the conjugate fast variable is still present in the evolution equations for the slow variables, the conjugate variable is expressed in terms of the remaining state and conjugate variables. The fast conjugate variable is found as the solution to the evolution equation of the fast state variable evaluated at the MaxEnt value of the fast state variable. The conjugate fast variable is thus determined by compatibility of the MaxEnt value of the fast variable and the evolution equation for the fast variable. This way we end up with a vector field for the slow variables (on manifold $\NNN$) compatible with the MaxEnt embedding of the slow manifold into the original manifold. The DynMaxEnt method gives similar results to the Chapman-Enskog asymptotic expansion. It suffers from the same drawback that irreversible evolution is typically needed also on $\MMM$ while removing the drawback of unknown form of the asymptotic expansion.

Recently, it was proposed that rather than obtaining the lower-level evolution via projection between state variables on the two levels, one might seek a reduction entailing directly a relation between the upper and lower vector fields \cite{PTRSA2020}. The essence is the replacement of the upper time evolution of the state variables with a corresponding time evolution of the vector field (of the fluxes). The fixed point in this flux time evolution is the vector field generating the reduced mesoscopic time evolution. In addition, a Poisson-Grad hierarchy was proposed as a new reformulation of general kinetic equations that couples kinetic theory with hydrodynamics while preserving the Hamiltonian kinematics of both theories. 

A versatile reduction method is the projection operator technique by Zwanzig, Mori and others, \cite{zwanzig,Zubarev,hco,Espanol-plateau}. The method is capable to produce irreversible evolution on $\NNN$ even from purely reversible evolution on $\MMM$, but it typically ends up with memory kernels (i.e. integro-differential equations) or non-symmetric dissipative brackets \cite{hco}.

Another robust method of projecting the vector field $\XX$ on $\YY$  was formulated by Bruce Turkington in \cite{Turkington}. The reduction consists of the following steps. Consider a manifold $\MMM$. Liouville equation for the probability distribution function on the manifold is formulated, and linear projection from the distribution function is defined, range of which determines a manifold $\NNN$. Shannon entropy is assumed for the distribution function, which forms and embedding $\upi$ of $\NNN$ onto $\MMM$.

Let us first project Hamiltonian mechanics on $\MMM$ (the Liouville equation) to Hamiltonian mechanics on $\NNN$. The upper\footnote{The more detailed level is referred to  as the upper while the less detailed (reduced) as lower.} Poisson bivector $\uLL$ is projected as a twice contravariant tensor field on the space of state variables and, if necessary, evaluated at the MaxEnt embedding,
\begin{equation}
	\dL^{ab} = \left(\frac{\partial \dpi^a}{\partial x^i} \uL^{ij}(x) \frac{\partial \dpi^a}{\partial x^i}\right)\Big|_{x=\upi(y)}.
\end{equation}
To construct the Hamiltonian vector field on the lower level one further needs a Hamiltonian, energy on the lower level.

Let energy on $\MMM$ be $\uE(x)$. Energy on the lower level $\NNN$ is inherited from the higher level through the MaxEnt mapping $\dE(y) = \uE(\upi(y))$. However, since some energy modes present on the higher level have already been damped on the lower level, typically $\dE(\dpi(x))\neq \uE(x)$. If the latter relation were an equality, one could project the higher-level evolution to the lower-level easily as one would obtain that time derivative of $\dpi(x)$ be equal to $\dLL \cdot d\dE$, which would be the lower-level purely Hamiltonian vector field. Since, however, the equality typically does not hold, simple projection does not give the desired result.

Instead, a lack-of-fit Lagrangian is defined which compares projections of the exact trajectories on $\MMM$ with trajectories on $\NNN$. Minimization of the Lagrangian then leads to a GENERIC evolution on $\NNN$ and gives a dissipation potential driving thermodynamic evolution on $\NNN$. 

The method was applied for instance on two-dimensional turbulence by taking a truncating Fourier-Galerkin expansion in \cite{Turkington-turbulence} and in \cite{Turkington-point-vortex} by considering the canonical Hamiltonian evolution of point vortices. Furthermore, the fluctuation-dissipation theorem \cite{callen} was interpreted within the context of the reduction method in \cite{Turkington-fluctuations}. The method was then generalized in \cite{Kleeman2015,Kleeman2019} using a path-integral approach so that a sequence of reductions is also a reduction. Still, however, the evolution on the detailed level seems to be restricted to Liouville equation for densities.

Novelty of this paper lies in the following points.
The method developed by Turkington in \cite{Turkington} is generalized so that the reduction can start with reversible and irreversible evolution (in the GENERIC form) on a detailed (upper) level of description (with arbitrary entropy, energy, Poisson bracket and dissipation potential) and end with GENERIC evolution on a less detailed (lower) level of description. This general reduction technique is then illustrated on a harmonic oscillator and on the reduction from kinetic theory to hydrodynamics and mechanical equilibrium.

\section{Static reduction}\label{sec.static}
Before discussing the dynamical reduction, let us first recall the principle of maximum entropy (MaxEnt) and its geometrical version via two consecutive Legendre transformations \cite{redext,PKG}.

\begin{figure}[ht!]
\begin{center}
 \begin{tikzpicture}[scale=1.0, every node/.style={scale=1.0}]
 \tikzset{level/.style={rounded corners, shape=rectangle, draw, fill=blue!10, font=\small}}
 \node[level](x){$\xx$};
 \node[level](x*)[right = 50pt of x]{$\xx^*$};
 \node[level](y)[below = 50pt of x]{$\yy$};
 \node[level](y*)[right = 50pt of y]{$\yy^*$};
 \begin{pgfonlayer}{bg}
   \draw[<->](x) to node [pos=0.5,above] {LT} (x*);
   \draw[<->](y) to node [pos=0.5,above] {LT} (y*);
   \draw[->](x) to node [pos=0.5,left] {$\dpi$} (y);
   \draw[->](y*) to node [pos=0.5,above] {MaxEnt} (x) ;
   \draw[->](y*) to node [pos=0.5,right] {$\upi$} (x*);
 \end{pgfonlayer}
\end{tikzpicture}
\caption{Relations between state variables on the higher level of description and the lower level and their conjugates. }
\label{fig.static}
\end{center}
\end{figure}
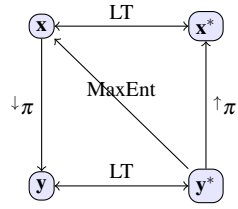

Consider an upper (more detailed) level of description with state variables $\xx\in\MMM$ and a lower (less detailed) level with variables $\yy\in\NNN$ such that the lower level state variables are given by a linear projection of the higher variables, i.e.
\begin{equation}\label{eq.projection}
 \yy = \dpi(\xx).
\end{equation}
The assumption of linearity (or at least affinity) of mapping $\dpi$ will be used later, especially in \ref{sec.MaxEnt}.\footnote{Note that nonlinear projections are also possible. For instance the mapping from the one-particle distribution function to the local-in-space entropy density
$$s(\rr) = -k_B \int\diff \pp f(\ln(h^3 f)-1)$$
is not linear as a function of $f$. However, we require linearity as it is necessary for the existence of derivatives of the projections used in the construction of the MaxEnt estimate of the upper detailed evolution, see Sec. \ref{sec.MaxEnt}. As a result, we can not project $f$ to entropy density, but we choose energy density, projection to which is linear.}
Fig. \ref{fig.static} shows relations among the variables and their conjugates. The relation between conjugate variables $\yy^*$ and $\xx^*$ is then given by the principle of maximum entropy (MaxEnt) and was summarized for example in \cite{redext}.
Details of the particular connections in the diagram can be found in Appendix \ref{sec.static.details}. Relations between lower, upper and conjugate energies can be found in \ref{sect.Projection}.

In particular, MaxEnt, i.e. maximization of entropy subject to given constraints, can be seen geometrically as a sequence of a generalized Legendre transformation followed by a standard Legendre transformation, see \ref{sec.MaxEnt}.

\section{Dynamic reduction}
Let us now show how to reduce GENERIC evolution on the upper level to evolution on the lower level.

\subsection{GENERIC on the upper level}\label{sec.GENERIC}
The higher-level evolution is assumed to be in the GENERIC form
\begin{equation}\label{eq.evo.xi}
\dot{x^i} = \uL^{ij} \frac{\partial \uE}{\partial x^j} + \frac{\partial \uXi}{\partial x^*_i}\Big|_{\xx^*=\uS_\xx},
\end{equation}
where the antisymmetric Poisson bivector $\uLL$ forms a Poisson bracket
\begin{equation}
 \uPB{F}{G} = \frac{\partial F}{\partial x^i} \uL^{ij} \frac{\partial G}{\partial x^j},
\end{equation}
satisfying Jacobi identity, and $\uXi$ is a dissipation potential. GENERIC evolution can be seen as constructed from the following building blocks: Once the state variables (determining the level of description or manifold of states) is declared, one needs a Poisson bivector (or a Poisson bracket), which is typically the same for all physical systems on the particular level of description, see e.g. \cite{PKG}. To  have the whole reversible part of the evolution, one needs an energy functional, which is already system-specific, and which is plugged into the Poisson bivector in order to create the reversible vector field (right hand side of the evolution equations). In order to add the irreversible evolution, an entropy functional is needed so that it can be used in gradient dynamics, where the vector field is given by derivatives of a dissipation potential with respect to conjugate variables (derivatives of entropy). In summary, for a given set of state variables the Poisson bracket is usually known, but we have to add system-dependent functionals of energy, entropy and a dissipation potential.

The GENERIC evolution splits into the reversible Hamiltonian part and irreversible gradient part. Gradient dynamics (also called generalized gradient dynamics in the case of non-quadratic $\uXi$) is a geometrical construction of dissipative (entropy growing) and irreversible (with respect to time-reversal transformation \cite{pre2014}) process. Such splitting has been observed also in the framework of natural configurations in continuum thermodynamics \cite{malek.j.prusa.v:derivation}. As being geometrical, gradient dynamics is invariant with respect to change of variables, see Appendix \ref{sec.GD}.
The dissipation potential is typically assumed to be convex in the conjugate variables, since the second law of thermodynamics is then implied, see e.g. Appendix \ref{sec.convex}, but it can also be non-convex without violating the second law \cite{nonconvex}. Gradient dynamics is in close relation with entropy production maximization \cite{Adam-EPM,rajagopal2004,Malek-Maxwell} and the steepest entropy ascent (SEA) \cite{SEA-GENERIC}.

Gradient dynamics can be also regarded as a consequence of fluctuations obeying a large deviations principle \cite{mielke2014,mielke-peletier}.

Since energy is conserved in GENERIC and entropy grows (assuming isolated systems), there is a natural Lyapunov functional driving the evolution towards thermodynamic equilibrium (exergy, available work or simply the thermodynamic potential, see e.g. \cite{PKG}). In the case of open systems, however, another potential plays the role of Lyapunov functional \cite{Vitek-Entropy}, and there is hope that applicability of the potential could be extended in the GENERIC framework.

From the geometric point of view the two parts of GENERIC, Hamiltonian and gradient, are two rather different concepts. For instance, the former is generated by an antisymmetric operator (Poisson bivector) while the latter by a symmetric one (a Riemannian metric or its non-linear generalization \cite{SEA-GENERIC}). A possible remedy for this dichotomy capturing both parts within one geometrical concept is the contact geometry \cite{arnoldbook,ent16,esen2012geometry,hermann,PKG}. We shall, however, follow the usual approach to GENERIC in the present work.

\subsection{Poisson brackets}
Let us first construct geometric objects that are necessary for the generalization of the lack-of-fit reduction method. We shall often need to consider state variables and their conjugates, since the latter generate evolution of the former. More details can be found in Sec. \ref{sec.static.details}. 

In the GENERIC upper evolution \eqref{eq.evo.xi} the reversible part is generated by a Poisson bivector. Motivated by \cite{Turkington}, we need to pass to the conjugate variables. The conjugate Poisson bivector on the higher level is then given by
\begin{equation}
 \uL^*_{ij} = \frac{\partial x^*_i}{\partial x^k}\uL^{kl}\frac{\partial x^*_j}{\partial x^l}
	= (d^2 \uS)_{ik} \uL^{kl} (d^2 \uS)_{lj}.
\end{equation}
Note that the differentiation is interpreted as functional derivatives, summation over repeated indexes can be integration, in which case the boundary terms are dropped (isolated or boundary-less systems, e.g. a torus), please discuss \cite{PKG} for more details.

We shall also need Hamiltonian dynamics on the lower level of description. The projected bivector on the lower level is given by
\begin{equation}
 \dL^{ab} = \frac{\partial y^a}{\partial x^i}\uL^{ij}\frac{\partial y^b}{\partial x^j},
\end{equation}
see  \cite{Hierarchy} for more details.
If the right hand side of this equation is closed (expressed only in terms of the lower-level variables, $\yy$), the bivector $\dLL$ is also Poisson and it generates a Poisson bracket (including Jacobi identity). If not, the expression has to be closed by evaluating the remaining higher-level variables in terms of the lower-level ones. The dependence $\xx(\yy^*(\yy))$ was derived in Sec. \ref{sec.static}, see equations \eqref{eq.uS.d.y} and \eqref{eq.y*.y}. This way we obtain a reduced lower-level bivector
\begin{equation}
 \dLr^{ab} = \dL^{ab}|_{\xx(\yy^*(\yy))},
\end{equation}
which is still antisymmetric but which may not fulfill Jacobi identity anymore. Conjugate reduced lower-level bivector is defined as
\begin{equation}\label{eq.dLr*}
 \dLr^*_{ab} = \frac{\partial y^*_a}{\partial y^c}\dLr^{cd}\frac{\partial y^*_b}{\partial y^d}
	= (d^2 \dS)_{ac} \dLr^{cd} (d^2 \dS)_{db}.
\end{equation}
The Poisson bracket on the lower level is then calculated as
\begin{equation}
	\dPB{F}{G} = \frac{\partial F}{\partial y^a} \dLr^{ab} \frac{\partial G}{\partial y^b}.
\end{equation}
Concrete examples follow in Sec. \ref{sec.app}. Therefore, action of the lower conjugate bivector can be expressed defining the lower conjugate Poisson bracket as
\begin{equation}
	\dPBc{F}{G}\stackrel{\mathrm{def}}{=}
	\frac{\partial F}{\partial y^*_a} \dLr^*_{ab} \frac{\partial G}{\partial y^*_b}
	=\frac{\partial F}{\partial y^*_a}
	\frac{\partial y^*_a}{\partial y^c} \dLr^{cd} \frac{\partial y^*_b}{\partial y^d}
	\frac{\partial G}{\partial y^*_b}
	=\frac{\partial F}{\partial y^c}
	 \dLr^{cd}
	\frac{\partial G}{\partial y^d} = \dPB{F}{G},
\end{equation}
which simplifies the calculations below.

See \ref{sec.static.details} for the relations between lower, conjugate and apparent energies.

\subsection{The lack-of-fit Lagrangian}

We shall seek a lack-of-fit Lagrangian that identifies a lower-level evolution entailing the best estimate (in a certain way specified below) of the actual evolution on the upper detailed level. To this end, let us first define a new upper-level energy, the \textit{apparent upper-level energy}, being the energy anticipated from observations of the lower-level evolution only:
\begin{equation}
	\uEb(\xx) \stackrel{def}{=} \dE(\dpi(\xx)) = \uE(\xx(\yy^*(\yy(\xx)))) \neq \uE(\xx).
\end{equation}
Naturally, its differential can be rewritten solely in terms of lower-level variables and functionals
\begin{equation}
	\frac{\partial \uEb}{\partial x^i} = \frac{\partial y^a}{\partial x^i}\frac{\partial \dE}{\partial y^a}\Big|_{\yy(\xx)}.
\end{equation}

Consider now the exact upper GENERIC evolution $\dot{x}^i$, given by Eq. \eqref{eq.evo.xi}. The purpose is to find such lower conjugate evolution $\dot{y}^*_a$ that the MaxEnt mapping from it best approximates the exact evolution. To formulate the difference between the exact upper evolution and the MaxEnt image of the approximate lower evolution, we construct the residuum (again motivated by \cite{Turkington})
\begin{equation}
	R^i = \dot{x}^i - \frac{\partial x^i}{\partial y^*_a}\dot{y}^*_a
	= \uL^{ij}\frac{\partial \uEb}{\partial x^j} + \frac{\partial \uXi}{\partial x^*_i} - \frac{\partial x^i}{\partial y^*_a}\dot{y}^*_a.
\end{equation}
Note the presence of apparent upper energy, $\uEb$ 
which, as we show below, is necessary to obtain lower conjugate bivector of the variation of action.

The lack-of-fit Lagrangian is defined using the entropic norm (see \cite{gorban}) of the residuum
\begin{equation}\label{eq.Lagr}
	\Lagr(\yy^*(t), \dyy^*(t)) = -\frac{1}{2}R^i(x(\yy^*)) \frac{\partial^2 \uS}{\partial x^i \partial x^j}\Big|_{\xx(\yy^*)} R^j(x(\yy^*)),
\end{equation}
where the MaxEnt values of $\xx$ were substituted into the expressions on the upper level of description. Note that the minus sign makes the Lagrangian positive definite, since the second differential of entropy is negative definite (concave function). 

For a given time $t_1$ and state $\yy^*_1$ we require the dissipation potential, which serves as the action, to follow from the end point optimization principle
\begin{equation}
	\Psi(t_0, \yy^*_0) = \underset{\yy^*(t_0)=\yy^*_0}{min} \int_{t_0}^{t_1} \diff t \, \Lagr(\yy^*(t), \dyy^*(t)).
\end{equation}
where $t_0$ is an initial time. The motivation for such a definition arises in the need to relax the assumption of the end point being an equilibrium. Thence we optimize over all (detailed) trajectories which at time $t_1$ yield the same state $\yy^*_1$. Additionally, this end state $\yy^*_1$ is not arbitrary, but rather being a point on the slow manifold which we assume to be an attractor of the detailed evolution (an important special case corresponds to equilibrium), c.f. \cite{Turkington}. As a result, we cannot rule out the dependence of this optimization procedure on the initial time $t_0$. However, we expect that this stationarity (independence of $\Psi$ on $t_0$) is reasonable when there is a time-scale separation between fast and slow processes where the former corresponds to a complete detailed evolution while the latter to the sought evolution on the lower level.

The extremal trajectory $\yy^*(t)$ and the action $\Psi$ can be sought by means of Hamilton-Jacobi theory
and they satisfy the following equations (we drop the lower index $0$ in the unknowns $(t_0,\yy_0^*)$ hereafter)\footnote{To obtain Hamilton-Jacobi equation for free starting point and fixed end point we substitute the general relation for variation, Eq (12) from Section 13 of \cite{Gelfand-Fomin}, with $\delta x_1=\delta y_1=0$ into Eq (69) in Section 23.}
\begin{subequations}
	\begin{eqnarray}
		\label{eq.HJ1} -\frac{\partial \Psi}{\partial y^*_a} &=& \frac{\partial \Lagr}{\partial \dy^*_a}\\
		\label{eq.HJ2}\frac{\partial \Psi}{\partial t} &=& \Hami\left(t, y^*_1, -\frac{\partial \Psi}{\partial y^*_a}\right),
	\end{eqnarray}
\end{subequations}
where $\Hami$ is the Legendre transformation of $\Lagr$,
\begin{equation}\label{eq.Hami}
 \Hami(\yy^*, \zz) = -\Lagr(\yy^*,\dyy^*(\yy^*,\zz)) + z^a \dy^*_a(\yy^*, \zz)
\end{equation}
evaluated at a solution (root) of
\begin{equation}\label{eq.Lagr.d}
 \frac{\partial}{\partial \dyy^*}\left(-\Lagr + z^a \dy^*_a\right) = 0, \mbox{  i.e. }
 z^a = \frac{\partial \Lagr}{\partial \dy^*_a}.
\end{equation}
Hamilton-Jacobi equation \eqref{eq.HJ2} is the analogue of the non-stationary optimization principle from \cite{Turkington}. An interesting feature of this evolutionary operator equation is that when one starts at a point where $d\Psi=0$, i.e. with no entropy production at the beginning, then it linearly increases in time, which is the so called plateau effect, see also \cite{Espanol-plateau}.

Assuming that the action be time-independent (seeking an universal law after the plateau effect), we obtain from Eq. \eqref{eq.HJ2} the stationary Hamilton-Jacobi equation
\begin{equation}
	\Hami\left(\yy^*, -\frac{\partial \Psi}{\partial \yy^*}\right) = 0,
\end{equation}y
which is an equation for $\Psi(\yy^*)$.

Using the concrete expression for the Lagrangian \eqref{eq.Lagr}, equation \eqref{eq.HJ1} becomes
\begin{eqnarray}
- \frac{\partial \Psi}{\partial y^*_a} &=&
	-R^i(\xx(\yy^*)) (\diff^2 \uS)_{ik}(\xx(\yy^*))\left(-\frac{\partial x^k}{\partial y^*_a}\right)\nonumber\\
	&=& 
	\left(\uL^{ij}\frac{\partial \uEb}{\partial x^j} + \frac{\partial \uXi}{\partial x^*_i} - \frac{\partial x^i}{\partial y^*_b}\dot{y}^*_b\right) \frac{\partial x^*_i}{\partial x^k}\Big|_{\xx(\yy^*)} \frac{\partial x^k}{\partial y^*_a}\nonumber\\
	&\stackrel{\mbox{Eq.\eqref{eq.A}, \eqref{eq.uS.d.y}}}{=}&
	-\frac{\partial^2 \dS^*}{\partial y^*_a \partial y^*_b} \dot{y}^*_b 
	+\frac{\partial y^*_c \pim{c}{i}}{\partial y^*_a} \uL^{ij}\frac{\partial \dE}{\partial y^b}\frac{\partial y^b}{\partial x^j}
	+\frac{\partial \uXi}{\partial x^*_i}\Big|_{\xx(\yy^*)} \frac{\partial x^*_i}{\partial y^*_a}\nonumber\\
	&=&
	-\dot{y}^a
	+ \dLr^{ae}\frac{\partial \dE}{\partial y^e}
	+\frac{\partial \uXi}{\partial x^*_i}\Big|_{\xx^*(\yy^*)}\frac{\partial x^*_i}{\partial y^*_a}.
\end{eqnarray}
Thus we arrive at
\begin{equation}\label{eq.GENERIC}
 \dy^a = \dLr^{ab}\frac{\partial \dE}{\partial y^b}
	+\frac{\partial}{\partial y^*_a}\underbrace{\left( \uXi(\xx(\yy),\xx^*(\yy^*)) + \Psi(\yy^*)\right)}_{=\dXi(\yy^*)},
\end{equation}
which can be referred to as preGENERIC, c.f. \cite{Kraaij2018}, evolution equation (preGENERIC instead of GENERIC, introduced in \cite{go,og}, because the bivector may not fulfill the Jacobi identity). If the Jacobi identity is fulfilled, we obtain reduced evolution equations that are fully compatible with GENERIC. The conjugate lower variables are identified with derivatives of lower entropy as in Eq. \eqref{eq.dS.d}. The lower dissipation potential $\dXi$ is identified as the sum of the upper dissipation potential evaluated at the MaxEnt mapping and the action at the extremal trajectory.

Evolution of the lower conjugate variable reads
\begin{equation}\label{eq.GENERIC*}
 \dy^*_a = \dLr^*_{ab}\frac{\partial \dE^*}{\partial y^*_b}
	+\frac{\partial^2 \dS}{\partial y^a\partial y^b} \frac{\partial \dXi}{\partial y^*_b},
\end{equation}
as results from multiplication of \eqref{eq.GENERIC} by $\frac{\partial y^*_b}{\partial y^a}$.

\subsection{Dissipation potential}
Let us now explicitly find the equation determining the dissipation potential $\dXi(\yy^*)$. Firstly, the residuum evaluated at the extremal trajectory become
	\begin{eqnarray}
		R^i(\yy, \dot{\yy}^*) 
		= \uL^{ij}\frac{\partial \uEb}{\partial x^j}\Big|_{\xx(\yy^*)} 
		+ \frac{\partial \uXi}{\partial x^*_i}\Big|_{\xx(\yy^*)} 
		- \frac{\partial x^i}{\partial y^*_a} \left(\dLr^*_{ab}\frac{\partial \dE^*}{\partial y^*_b}
		+\frac{\partial^2 \dS}{\partial y^a\partial y^b} \frac{\partial \dXi}{\partial y^*_b}\right).
	\end{eqnarray}
The stationary Hamiltonian-Jacobi equation then reads
\begin{eqnarray}
	0 &=& -\frac{1}{2} R^i (\diff^2 \uS)_{ij} R^j + \frac{\partial \Psi}{\partial y^*_a}
	\left(
  	\dLr^*_{ab}\frac{\partial \dE^*}{\partial y^*_b}
	+\frac{\partial^2 \dS}{\partial y^a\partial y^b} \frac{\partial \dXi}{\partial y^*_b}
	\right)
\end{eqnarray}
from which the action $\Psi$ can be determined once entropy $\dS$, lower Poisson bivector $\dLr$ and upper dissipation potential $\uXi$ are specified. Using for instance relation \eqref{eq.uS.d.y}, the Hamilton-Jacobi equation can be simplified to 
\begin{multline}\label{eq.HJ}
-\frac{1}{2}\left(\uLL\cdot \diff \uEb|_{\xx(\yy^*)}-\dPBc{\xx(\yy^*)}{\dE^*}\right)^T 
	\cdot \diff^2 \uS \cdot
	\left(\uLL\cdot \diff \uEb-\dPBc{\xx(\yy^*)}{\dE^*}\right)\Big|_{\xx(\yy^*)}\\
	= -\dPBc{\Psi}{\dE^*} 
	-\frac{1}{2}\frac{\partial \uXi}{\partial \xx^*} \cdot\left(\diff^2 \uS- \ppi^T\diff^2 \dS\cdot\ppi\right) \cdot \frac{\partial \uXi}{\partial \xx^*} 
	-\frac{1}{2}\frac{\partial \Psi}{\partial \yy^*} \cdot \diff^2 \dS \cdot \frac{\partial \Psi}{\partial \yy^*}\\
	-\frac{1}{2}\left(\frac{\partial \uXi}{\partial \yy^*} \diff^2 \dS \frac{\partial \Psi}{\partial \yy^*}
	+\frac{\partial \Psi}{\partial \yy^*} \diff^2 \dS \frac{\partial \uXi}{\partial \yy^*}\right)\\
	+\frac{1}{2}\left(\frac{\partial \uXi}{\partial \xx^*} \left(\diff^2 \uS -\ppi^T \cdot \diff^2 \dS \cdot \ppi\right)\cdot \uLL\cdot \frac{\partial \uEb}{\partial \xx} +(\dots)^T \right),
\end{multline}
where $\dots$ denote the same expression as preceding.

If there is no upper dissipation potential, the Hamilton-Jacobi equation becomes
\begin{multline}\label{eq.HJ.noXi}
-\frac{1}{2}\left(\uLL\cdot \diff \uEb|_{\xx(\yy^*)}-\dPBc{\xx(\yy^*)}{\dE^*}\right)^T 
	\cdot \diff^2 \uS \cdot
	\left(\uLL\cdot \diff \uEb-\dPBc{\xx(\yy^*)}{\dE^*}\right)\Big|_{\xx(\yy^*)}\\
	= -\dPBc{\Psi}{\dE^*} 
	-\frac{1}{2}\frac{\partial \Psi}{\partial \yy^*} \cdot \diff^2 \dS \cdot \frac{\partial \Psi}{\partial \yy^*},
\end{multline}
which can be simplified to the Riccati equation \eqref{eq.Riccati} if the Poisson bivector is constant and if the action is quadratic, as we shall see below. In this case we have $\Psi = \dXi$.

Moreover, if the projection of the Poisson bivector disappears (no Hamiltonian evolution on the lower level), then the Hamilton-Jacobi equation further simplifies to 
\begin{multline}\label{eq.HJ.noXi.noL}
-\frac{1}{2}\left(\uLL\cdot \diff \uEb|_{\xx(\yy^*)}\right)^T 
	\cdot \diff^2 \uS \cdot
	\left(\uLL\cdot \diff \uEb\right)\Big|_{\xx(\yy^*)}
	=
	-\frac{1}{2}\frac{\partial \Psi}{\partial \yy^*} \cdot \diff^2 \dS \cdot \frac{\partial \Psi}{\partial \yy^*},
\end{multline}
which is a quadratic equation for the functional $\Psi$. If a solution to the Hamilton-Jacobi equation exists, then the lower evolution is dissipative although the upper evolution was purely Hamiltonian, see also Sec. \ref{sec.LHO}.

In summary, the reduced evolution on the lower level is determined by GENERIC evolution on the upper level, projection $\dpi$ and entropy. The above method provides a tool how to identify a consistent dissipation with the detailed evolution, however, a linear mapping between the two connected levels is required. This restriction is crucial as it is necessary for the existence of derivatives of the projections used in the construction of the MaxEnt estimate of the upper detailed evolution. Note, however, that although nonlinear projections are standardly used between levels, there might be linear projections at hand  even though we usually consider non-linear evolution  as we demonstrate in the Boltzmann to Navier-Stokes reduction, where the nonlinear projection (in the state variable, i.e. in the distribution function $f$) to entropy is replaced by a linear projection to energy.

\subsection{The quadratic case}
Let us first formulate a case where all potentials are quadratic and Poisson bivectors are constant (canonical).

\subsubsection{Entropies}
State variables are denoted again by $\xx$ on an upper level while $\yy$ on a lower level.

Let entropy on the upper level be
\begin{equation}
	\uS = \frac{1}{2}(d^2 \uS)_{ij}x^i x^j,
\end{equation}
where $d^2\uS$ is the negative definite operator of second differential of the (concave) upper entropy.
The conjugate entropy is then
\begin{equation}
 \uS^* = \frac{1}{2}(d^2 \uS^*)^{ij} x^*_i x^*_j,
\end{equation}
where $d^2\uS^*$ is the inverse operator to $d^2 \uS$ and is also negative definite.

Let the projection to the lower level be given by a linear mapping (e.g. a matrix) as
\begin{equation}
	y^a = \pim{a}{i} x^i.
\end{equation}
The conjugate entropy on the lower level is then
\begin{equation}
	\dS^* = \frac{1}{2}\underbrace{(d^2 \uS^*)^{ij}\pim{a}{i} \pim{b}{j}}_{=(d^2 \dS^*)^{ab}} y^*_a y^*_b,
\end{equation}
where the (also negative definite) operator $d^2 \dS^*$ was introduced.
Finally, entropy on the lower level is
\begin{equation}
 \dS = \frac{1}{2} d^2 \dS_{ab}y^a y^b,
\end{equation}
where $d^2 \dS$ is the inverse operator to $d^2 \dS^*$.

When the upper entropy is algebraic (not dependent on spatial derivatives of state variables), i.e. $d^2 \uS$ is a matrix, then all the other operators and entropies can be calculated easily using matrices.

\subsubsection{Energies}
Assuming that the upper energy is quadratic as well,
\begin{equation}
	\uE = \frac{1}{2}(d^2 \uE)_{ij} x^i x^j,
\end{equation}
the upper conjugate energy becomes
\begin{equation}
	\uE^*(\xx^*) = \uE(\xx(\xx^*)) = \frac{1}{2}\underbrace{(d^2 \uE)_{ij} (d^2 \uS^*)^{ik} (d^2 \uS^*)^{jl}}_{=(d^2 \uE^*)^{kl}} x^*_k x^*_l.
\end{equation}

The lower conjugate energy is then
\begin{equation}
	\dE^*(\yy^*) = \frac{1}{2}\underbrace{\pim{a}{l}(d^2\uS^*)^{lj} (d^2 \uE)_{jk}\pim{b}{i}(d^2 \uS^*)^{ki}}_{=(d^2 \dE^*)^{ab}} y^*_a y^*_b,
\end{equation}
and the lower energy
\begin{equation}
	\dE(\yy) = \frac{1}{2} \underbrace{(d^2 \dE^*)^{ab} (d^2 \dS)_{ac} (d^2 \dS)_{bd}}_{(d^2 \dE)_{cd}} y^c y^d.
\end{equation}
The lower energy and lower conjugate energy are thus also quadratic.

\subsubsection{Poisson bivector}
We assume that the Poisson bivector on the upper level is constant (as for instance in particle mechanics or smooth particle hydrodynamics \cite{Ellero}.
Having a constant Poisson bivector $\uL^{ij}$, the lower-level Poisson bivector becomes
\begin{equation}
	\dL^{ab} = \pim{i}{a}\uL^{ij}\pim{j}{b},
\end{equation}
and its conjugate
\begin{equation}
	\dL^*_{cd} = (d^2 \dS)_{ca} \dL^{ab} (d^2 \dS)_{db}.
\end{equation}
Note that the lower-level bivector is antisymmetric and constant, which means that it surely fulfills Jacobi identity, see e.g. \cite{hco}.

\subsubsection{Quadratic dissipation potential}
A quadratic dissipation potential $\uXi$ on the upper level of description can be expressed as
\begin{equation}
 \uXi = \frac{1}{2} \uM^{ij} x^*_i x^*_j,
\end{equation}
where $\uM$ is a symmetric positive semi-definite operator, e.g. a dissipative matrix, on the upper level.

From the structure of the Hamilton-Jacobi equation, the structure of $\Psi$ can be also assumed quadratic as will be apparent below
\begin{equation}\label{eq.Psi.quad}
 \Psi = \frac{1}{2} y^*_a \psi^{ab} y^*_b
\end{equation}
where $\psi^{ab}$ is again a symmetric positive semi-definite operator.

Hamilton-Jacobi equation \eqref{eq.HJ} becomes\footnote{We require the equation to be satisfied identically, i.e. for all values of the lower state variables. Given the assumptions on the structure of the dissipation potential, we require all the coefficients of $y^*_a y^*_b$ to vanish.}
\begin{multline}\label{eq.HJ.quadratic}
	\frac{1}{2}\PPhi + \frac{1}{2} \dMM^T  \cdot\left(\diff^2 \uS - \ppi^T \cdot \diff^2 \dS \cdot\ppi\right) \cdot\dM  \\
	= -\frac{1}{2}\left( \psi^T  \cdot\dLL^*  \cdot\diff^2 \dE^* - (\diff^2\dE^*)^T  \cdot\dLL^*  \cdot \ppsi\right)
	-\frac{1}{2} \ppsi^T  \cdot\diff^2 \dS \cdot\ppsi,
\end{multline}
which is an operator equation for the unknown $\psi^{ab}$ symmetric operator. The lower dissipative matrix $\dMM$ is defined as
\begin{equation}
	\dM^{ab} = \pim{a}{j}\uM^{ij}\pim{b}{j},
\end{equation}
and the $\PPhi$ operator comes from the left hand side of the Hamilton-Jacobi equation \eqref{eq.HJ},
\begin{multline}\label{eq.Riccati}
	\frac{1}{2}y^*_c \Phi^{cd} y^*_d  =\\
	=-\frac{1}{2}\left(\uLL\cdot \diff \uEb|_{\xx(\yy^*)}-\dPBc{\xx(\yy^*)}{\dE^*}\right)^T 
	\cdot \diff^2 \uS \cdot
	\left(\uLL\cdot \diff \uEb-\dPBc{\xx(\yy^*)}{\dE^*}\right)\Big|_{\xx(\yy^*)}.
\end{multline}
If all the other terms are algebraic (matrices), then the quadratic Hamilton-Jacobi equation \eqref{eq.HJ.quadratic} becomes an algebraic Riccati equation for matrix $\psi^{ab}$,  see e.g. \cite{Kucera-Ricatti}.

Finally, let us mention a special case when there is no dissipation on the upper level, i.e. $\uXi=0$ and $\uM=0$, and no Poisson bivector on the lower level, i.e. $\dL=0$. The Hamilton-Jacobi equation becomes
\begin{equation}
	\PPhi = -\ppsi^T \cdot \diff^2 \dS \cdot \ppsi,
\end{equation}
which is a quadratic equation for matrix $\psi^{ab}$. Hence in a situation where the upper evolution is purely Hamiltonian and the lower is purely dissipative, the lack of fit matrix $\PPhi$ thus yields dissipation potential $\dXi=\Psi$, which drives the lower evolution towards equilibrium. 

\section{Applications}\label{sec.app}
Let us demonstrate the lack-of-fit reduction first on a an example of a harmonic oscillator. Then we show how kinetic theory reduces to hydrodynamics.

\subsection{Explicit example: a harmonic oscillator}\label{sec.LHO}
We shall now consider a harmonic oscillator. The evolution is reversible, but when we are regarding only the position of the oscillator as a state variable, the reduced evolution becomes purely irreversible, resembling friction.

\subsubsection{Reversible evolution on the upper level}
State of the oscillator can be described by the position $q$, momentum $p$ and entropy $s$. Energy of the oscillator is 
\begin{equation}
	E = \frac{p^2}{2m} + \frac{1}{2}\alpha q^2 + \epsilon(S),
\end{equation}
where the last term stands for internal energy of the oscillator. In the energetic representation the evolution equations governing evolution of the state variables, which are generated by the energy, are
\begin{subequations}\label{eq.evo.qps}
\begin{eqnarray}
	\dot{q} &=& E_p = \frac{p}{m}\\
	\dot{p} &=& -E_q = -\alpha q\\
	\dot{S} &=& 0.
\end{eqnarray}
Together they imply that energy is conserved,
\begin{equation}
	\dot{E} = 0.
\end{equation}
\end{subequations}
Equations \eqref{eq.evo.qps} are Hamiltonian evolution generated by Poisson bivector
\begin{equation}
	\uLL  = 
	\begin{pmatrix}
		0 & 1 & 0\\
		-1 & 0 & 0\\
		0 & 0 & 0
	\end{pmatrix},
\end{equation}
which is clearly degenerate (has a Casimir invariant) due to the presence of entropy. The Poisson bracket is canonical,
\begin{equation}
	\uPB{F}{G}=\left(\frac{\partial F}{\partial q}\right)_{p,S}\left(\frac{\partial G}{\partial p}\right)_{q,S}
-\left(\frac{\partial G}{\partial q}\right)_{p,S}\left(\frac{\partial F}{\partial p}\right)_{q,S}.
\end{equation}

In order to be able to formulate MaxEnt properly, we need concave entropy. But since entropy of the oscillator in the energetic representation is linear, being equal to $S$, we have to switch to the entropic representation, where energy $E$ is among the state variables, $\xx = (q,p,E)$. Plugging functionals dependent on $q,p,E$ into the canonical Poisson bracket, the bracket transforms to
\begin{align}\label{eq.PB.osc}
	\uPB{F}{G}=
	&
	\left(\frac{\partial F}{\partial q}\right)_{p,E}\left(\frac{\partial G}{\partial p}\right)_{q,E}
	-\left(\frac{\partial G}{\partial q}\right)_{p,E}\left(\frac{\partial F}{\partial p}\right)_{q,E}\nonumber\\
	&-\frac{\left(\frac{\partial S}{\partial p}\right)_{q,E}}{\left(\frac{\partial S}{\partial E}\right)_{q,p}}
	\left(\left(\frac{\partial F}{\partial q}\right)_{p,E}\left(\frac{\partial G}{\partial E}\right)_{q,p}
	-\left(\frac{\partial G}{\partial q}\right)_{p,E}\left(\frac{\partial F}{\partial E}\right)_{q,p}\right).
\end{align}
The generated evolution equations of course remain the same (when a concrete form of energy is employed). Entropy is now a concave function of internal energy, 
\begin{equation}
	\uS(q,p,E) = S\left(E-\frac{p^2}{2m} - \frac{1}{2}\alpha q^2\right),
\end{equation}
its differential is
\begin{equation}
	\diff \uS = S_E \left(-\alpha q, -\frac{p}{m}, 1\right),
\end{equation}
where the derivative $S_E = T^{-1}$ stands for inverse temperature.
The second differential is 
\begin{equation}
	\diff^2 \uS =
	\begin{pmatrix}
		S_{EE} \alpha^2 q^2 - S_E \alpha\quad &\quad S_{EE} \alpha q p/m\quad &\quad  -\alpha q S_{EE}\\
		S_{EE} \alpha q p/m & S_{EE} \frac{p^2}{m^2} - \frac{S_E}{m} & -S_{EE} \frac{p}{m}\\
		-\alpha q S_{EE} & -S_{EE} \frac{p}{m} & S_{EE},
	\end{pmatrix}
\end{equation}
where $S_{EE}$ is the second derivative of entropy with respect to energy (negative for concave entropies). 

In order to perform the following calculations explicitly, we have to choose a concrete dependence of entropy on the internal energy, for instance
\begin{equation}
	\uS(q,p,E) = k_B \ln \left(e-\frac{p^2}{2m} -\frac{1}{2}\alpha q^2\right) + c,
\end{equation}
where $c$ is a normalization constant making the expression inside logarithm dimensionless. This entropy has positive derivative with respect to energy (positive temperature) as well as positive second derivative with respect to energy.

\subsubsection{Static reduction}
Let us now regard the oscillator from a lower level of description where the state variables are represented by position and energy only, $\yy=(q,E)$. The projection operator $\ppi$ is thus given by
\begin{equation}
	\begin{pmatrix}
		q\\
		E
	\end{pmatrix}
	 = 
	\underbrace{
		\begin{pmatrix}
			1 & 0 & 0\\
			0 & 0 & 1
		\end{pmatrix}
	}_{=\ppi}
	\cdot
	\begin{pmatrix}
		q\\
		p\\
		E
	\end{pmatrix}
\end{equation}
and it is represented by a constant matrix.

Conversely, having position and energy, what is the MaxEnt (least biased) estimate of the momentum $p$? The reducing Legendre transformation (first step of MaxEnt) reads
\begin{subequations}
\begin{align}
	0 =& \frac{\partial}{\partial q}\left(-\uS + q^* q + E^* E\right) \\
	0 =&\frac{\partial}{\partial p}\left(-\uS + q^* q + E^* E\right)\\
	0 =&\frac{\partial}{\partial E}\left(-\uS + q^* q + E^* E\right).
\end{align}
\end{subequations}
Solution of these equations is
\begin{equation}
	q = -\frac{1}{\alpha} \frac{q^*}{E^*}, \qquad
	p = 0, \qquad
	E = \frac{k_B}{E^*} + \frac{1}{2\alpha} \frac{(q^*)^2}{(E^*)^2},
\end{equation}
which is the MaxEnt estimate once the lower conjugate variables are known.

The energy on the lower level is 
\begin{equation}
	\dE(q,E) = E
\end{equation}
and the upper apparent energy is 
\begin{equation}
	\uEb(q,p,E) = E .
\end{equation}

The conjugate lower entropy is then obtained by plugging the MaxEnt estimates into $-\uS + q^* q + E^* E$,
\begin{equation}
	\dS^*(q^*,E^*) = -k_B \ln \frac{k_B}{E^*}  -c + k_B - \frac{1}{2\alpha}\frac{(q^*)^2}{E^*}.
\end{equation}
Legendre transformation of the lower conjugate entropy is the lower entropy $\dS$,
\begin{equation}
	\dS(q,E) = k_B \ln\left(E-\frac{1}{2}\alpha q^2\right).
\end{equation}
Second differential of the lower entropy is
\begin{equation}
	\diff^2 \dS = 
	\begin{pmatrix}
		-\alpha \frac{k_B}{E-\frac{1}{2}\alpha q^2} -\frac{\alpha^2 q^2 k_B}{\left(E-\frac{1}{2}\alpha q^2\right)^2} \quad & \quad \frac{\alpha q k_B}{\left(E-\frac{1}{2}\alpha q^2\right)^2}\\
		 \frac{\alpha q k_B}{\left(E-\frac{1}{2}\alpha q^2\right)^2} & - \frac{k_B}{\left(E-\frac{1}{2}\alpha q^2\right)^2}
	\end{pmatrix}.
\end{equation}
So far we have finished the static reduction from $\xx=(q,p,E)$ to $\yy=(q,E)$. Let us now turn to the dynamic reduction.

\subsubsection{Dynamic reduction}
We did not consider any dissipation (no dissipation potential) on the upper level. 
Moreover, when functionals dependent only on $q$ and $E$ are plugged into the Poisson bracket and when the MaxEnt estimate $p=0$ is used, the Poisson bracket \eqref{eq.PB.osc} disappears, which means that there is not Poisson bivector on the lower level, $\dLLr =0$. The Hamilton-Jacobi equation \eqref{eq.HJ} thus simplifies to 
\begin{equation}\label{eq.HJ.osc}
	\frac{1}{2}\left(\uLL\cdot \diff \uEb|_{\xx(\yy^*)}\right)^T 
	\cdot \diff^2 \uS \cdot
	\left(\uLL\cdot \diff \uEb\right)\Big|_{\xx(\yy^*)}
	= 
	\frac{1}{2} \frac{\partial \Psi}{\partial \yy^*} \cdot \diff^2 \dS \cdot \frac{\partial \Psi}{\partial \yy^*}.
\end{equation}
The left hand side of this equation can be explicitly written as
\begin{align}
	\uL^{ij} \frac{\partial \uEb}{\partial x^j} (\diff^2 \uS)_{ik} \uL^{kl}\frac{\partial \uEb}{\partial x^l} 
	= -S_e \frac{\alpha^2 q^2}{2m} = -\frac{T}{2m} (q^*)^2.
\end{align}
Note that here we have replaced $E^*$ with $T^{-1}$ in order to get an expression quadratic in the conjugate variables. 
The right hand side can be sought in quadratic form \eqref{eq.Psi.quad}, assuming that only the $\Psi^{qq}$ entry be non-zero, leading to
\begin{equation}
	\Psi^{qq} = \frac{T}{m} \frac{\epsilon^2}{k_B} \frac{1}{\alpha(\epsilon + \alpha q^2)}
\end{equation}
for $\epsilon = E - \frac{1}{2}\alpha q^2$.
Evolution equation of the position is then given by Eq. \eqref{eq.GENERIC}, 
\begin{equation}
	\dot{q} = \Psi^{qq} q^* = -\frac{k_B T^2}{m(k_B T+\alpha q^2)} q.
\end{equation}
Therefore, a dissipative evolution equation for $q$ is obtained by the dynamical reduction of the completely reversible evolution on the upper level of description.

\subsection{Boltzmann $\rightarrow$ Navier-Stokes}

We shall apply the above method on a classical example of multilevel physics. In particular,  a relation between Navier-Stokes description of hydrodynamics and a micro (upper) level corresponding to description of ideal gas via reversible Vlasov equation. The upper level is characterized by state variable $f(t,\rr,\pp)$ (the one-particle distribution function) while the lower level is obtained by projection to the density ($m$ being mass of one particle), momentum density (per volume) and total energy density (per volume) of ideal gas
\begin{align}
  \label{eq:2}
  \rho(\rr) &= \int d \pp m f(\rr,\pp),\\
  u_i(\rr) &= \int d \pp p_i f(\rr,\pp),\\
  e(\rr) &= \int d \pp \frac{\pp^2}{2m} f(\rr,\pp).
\end{align}
Note in particular that the projection is chosen to the entropic representation (with energy as a state variable) rather than the projection to the energetic representation \cite{callen,PKG} due to the requirement of linearity of the projection.

\subsubsection{Hamiltonian mechanics}
The Poisson bracket describing reversible part of Vlasov equation is \cite{g1}
\begin{equation}\label{eq.PB.B}
  \{F,G\}^{(B)}=\int d \rr \int d \pp f(\rr,\pp) \left(\pd{r^i}F_f \pd{p_i} G_f- \pd{r^i}G_f \pd{p_i} F_f\right),
\end{equation}
and the implied Vlasov equation is
\begin{equation}
	\partial_t f = -\frac{\partial \uE_f}{\partial p_i} \frac{\partial f}{\partial r^i} +\frac{\partial \uE_f}{\partial r^i} \frac{\partial f}{\partial p_i}
\end{equation}
for arbitrary energy functional $\uE(f)$.

We shall now project it to the hydrodynamic level of description, where fields \eqref{eq:2} play the role of state variables.
Derivative of an arbitrary functional $A(\rho(f),\uu(f),e(f))$ can be rewritten as
\begin{eqnarray*}
	\pdv[A]{f(\rr,\pp)} &=&\int d \rr' \pdv[A]{\rho(\rr')} \pdv[\rho(\rr')]{f(\rr,\pp)}+
  \pdv[A]{u_i(\rr')} \pdv[u_i(\rr')]{f(\rr,\pp)}+
  \pdv[A]{e(\rr')} \pdv[e(\rr')]{f(\rr,\pp)}\nonumber\\
	&=&m A_\rho+p_i A_{u_i} + \frac{\pp^2}{2m} A_e.
\end{eqnarray*}
Poisson bracket \eqref{eq.PB.B} then becomes
\begin{align}
  \{A,B\}^{(B)} &= \int d \rr \int d \pp f (m \partial_k A_\rho+p_i \partial_k A_{u_i}) (B_{u_k}+\frac{p_k}{m} B_e)\nonumber\\
	&+\int d \rr \int d \pp f \frac{\pp^2}{2m} \partial_k A_e (B_{u_k}+\frac{p_k}{m} B_e)-\ldots=\nonumber\\
                &= \int d \rr \rho(\partial_k A_\rho B_{u_k}-\ldots) + u_i(\partial_k A_{u_i} B_{u_k}-\ldots)\nonumber\\
		&+\int d \rr u_k(\partial_k A_\rho B_e-\ldots)+e(\partial_k A_eB_{u_k}-\ldots) \nonumber\\
                &+ \int d \rr \left[\partial_k A_{u_i} B_e \underbrace{\int d \pp \frac{p_i p_k}{m} f}_{I} +\partial_k A_e B_e \underbrace{\int d \pp f \frac{\pp^2}{2m} \frac{ p_k}{m}}_{II}\right].
\end{align}
The last two terms cannot be fully expressed in the lower state variables and hence the Poisson bracket on the lower level in the entropic representation is not in general available (cf. with the energetic representation where the Poisson bracket is at hand \cite{PKG}).

To proceed we use the MaxEnt value of the distribution function corresponding to this projection,
\begin{subequations}
\begin{align}
  f(\rho,\uu,e) &= \frac{\rho}{m} \left(\frac{e^*}{2 \pi m k_B}\right)^{3/2} \exp\left(-\frac{m e^*}{2 k_B} \left(\frac{\uu}{\rho}-\frac{\pp}{m}\right)^2\right),\\
  e^* &= \frac{3}{2} k_B \frac{\rho}{m\left(e-\frac{\uu^2}{2 \rho}\right)},
\end{align}
\end{subequations}
which allows us to compute
\begin{subequations}
\begin{align}
  I &= \frac{u_i u_k}{\rho} + \delta_{ik} \frac{k_B \rho}{e^* m} =  \frac{u_i u_k}{\rho} + \delta_{ik} \frac{2}{3} \left(e-\frac{\uu^2}{2 \rho}\right),\\
  II &=\frac{1}{2} u_k \left(\frac{5 k_B}{e^* m} + \left(\frac{\uu^2}{\rho}\right)^2\right)=\frac{5}{3} \frac{u_k}{\rho} \left(e-\frac{\uu^2}{2\rho}\right)+\frac{1}{2} u_k \left(\frac{\uu}{\rho}\right)^2,
\end{align}
\end{subequations}
see e.g. \cite{PKG} for more details.

Using the shorthand notation of internal energy density $\epsilon=e-\frac{\uu^2}{2\rho}$, the projection of the upper Boltzmann Poisson bracket to the lower level of fluid mechanics in the entropic representation reads
\begin{eqnarray}
  \label{eq.NSE_PB}
	\{A,B\}^{(NS_E)} &=& \{A,B\}^{(B)}|_{f(\rho,\uu,e)} \\
	&=& \int d \rr \rho(\partial_k A_\rho B_{u_k} - \ldots) + u_i (\partial_i A_\rho B_e + \partial_k A_{u_i} B_{u_k}-\ldots) +\nonumber\\
	&&+\left(\frac{u_i u_k}{\rho} + \delta_{ik} \frac{2}{3} \epsilon\right)( \partial_k A_{u_i} B_e -\ldots) + e(\partial_k A_e B_{u_k}-\ldots)\nonumber\\
	&&+ \left(\frac{5}{3} \frac{u_k}{\rho} \epsilon + \frac{1}{u_k}\frac{\uu^2}{\rho^2}\right)(\partial_k A_e B_e - \ldots).\nonumber
\end{eqnarray}
This bracket is equivalent to the Poisson bracket for fluid mechanics in the energetic representation, i.e. in state variables $(\rho,\uu,s)$, transformed to the entropic representation using the energy of ideal gas. Therefore, it fulfills Jacobi identity and is indeed a Poisson bracket.

The lower conjugate Poisson bivector $\dL^*$ can be obtained from the direct lower Poisson bivector $\dL$, Eq. \eqref{eq.dLr*}, whose entries can be found in \cite{hco} or \cite{PKG}.
The hydrodynamic fields $(\rho,\uu,e)$ thus evolve in accordance with the Euler equations for compressible fluids combined with gradient dynamics,
\begin{equation}
	\partial_t
	\begin{pmatrix}
		\rho\\
		\uu\\
		e
	\end{pmatrix}
	= \mathrm{Euler} 
	+ \begin{pmatrix}
		\frac{\delta \Psi}{\delta \rho^*}\\
		\frac{\delta \Psi}{\delta \uu^*}\\
		\frac{\delta \Psi}{\delta e^*}
	\end{pmatrix},
\end{equation}
where the dissipation potential $\Psi$ is to be determined as a solution of the Hamilton-Jacobi equation.

\subsubsection{Dissipation potential}
We now aim to apply the above reduction technique to identify lower dissipation potential. To this end we need to find the lower conjugate entropy and its second differential, lower conjugate bivector (which is readily available from the above lower Poisson bracket) and the gradient of the lower energy. With our choice of state variables, where we have the energy as a state variable, the energy gradient is  $\pdv[E]{(\rho,\uu,e)} = (0,0,1)^T$.

The upper entropy is the Boltzmann entropy
\begin{equation*}
	\uS(f) = -\int \diff \rr \int \pp k_B f (\ln (h^3 f)-1),
\end{equation*}
and hence the lower entropy is the local equilibrium Sackur-Tetrode relation
\begin{equation*}
  \dS = \int d\rr k_B \frac{\rho}{m} \left(\frac{5}{2}+\ln\left[\frac{m}{\rho}\left(\frac{4 \pi m}{3 h^2}\frac{\epsilon}{\rho/m}\right)^{3/2}\right]\right).
\end{equation*}
Using results from \cite{PKG}, the conjugate lower entropy becomes
\begin{equation*}
	\dS^* = -\frac{k_B}{h^3} \int d\rr e^{-m\rho^*(\rr)/k_B} \left(\frac{2\pi m k_B}{e^*}\right)^{3/2} e^{\frac{1}{2} m \frac{(\uu^*)^2}{k_B e^*}}.
\end{equation*}

We are now ready to inspect the Hamilton-Jacobi relation for the dissipation potential on the lower level \eqref{eq.HJ}. Consider the case when there is no dissipation on the upper level (i.e. starting from ideal gas). Then the Hamilton-Jacobi equation for the lower dissipation potential has two terms, one being quadratic in its gradient while the second is linear in the gradient. Recalling the dependence of the second differential of the lower conjugate entropy (being an algebraic function of the lower state variables) and of the lower conjugate Poisson bivector (being linearly dependent on gradients of state variables), we observe that the gradient of the lower dissipation potential would have to be a linear function of gradient of state variables. This is somewhat unexpected as typically the dissipation appears via second order (Laplacian) terms. Hence we conclude that the reduction to Navier-Stokes requires a non-trivial dissipation functional on the upper level and we shall thus investigate a simpler example, the reduction to mechanical equilibrium.

\subsubsection{Mechanical equilibrium}
Let us proceed further by reducing to mechanical equilibrium, $\uu=0$, hence to the level with state variables $\rho,~e$.

From the lower Poisson bracket \eqref{eq.NSE_PB} we observe that the lower Poisson bivector vanishes (no reversible evolution) and the second differential of the lower conjugate entropy simplifies to
\begin{equation}
	d^2 \dS^* = \delta(\rr-\rr') e^{-m \rho^*(\rr')/k_B} \left(\frac{2 \pi m k_B}{h^2 e^*}\right)^{3/2}
	\begin{pmatrix}
		-\frac{m^2}{k_B} & -\frac{3}{2}\frac{m}{e^*}\\
		-\frac{3}{2}\frac{m}{e^*} & -\frac{15}{4}\frac{k_B}{(e^*)^2}
	\end{pmatrix},
\end{equation}
and the second differential of the lower entropy reads
\begin{equation}
	d^2 \dS = \delta(\rr-\rr')
	\begin{pmatrix}
		-\frac{5}{2}\frac{k_B}{m\rho} & \frac{3}{2} \frac{k_B}{m e}\\
		\frac{3}{2} \frac{k_B}{m e} & -\frac{3}{2} \frac{k_B \rho}{e^2 m}
	\end{pmatrix}.
\end{equation}
Determinant of the matrix is positive, which means that the matrix is negative definite as expected.

The Hamilton-Jacobi equation \eqref{eq.HJ.noXi.noL} becomes
\begin{align}
	\frac{1}{2k_B} &\int\diff\rr\left[ e^{-\frac{m \rho^*}{k_B}} \left(\frac{k_B m}{e^*}\right)^{5/2} \frac{\pi^{3/2}}{\sqrt{2}h^3 (e^*)^2}\cdot\right.\nonumber\\
	&\left.\cdot\left(35 k_B^2 m^2 (\nabla e^*)^2 + 20 e^* k_B m \nabla \rho^* \cdot \nabla e^* + 4(e^*)^2 (\nabla\rho^*)^2\right)\vphantom{\left(\frac{k_B m}{e^*}\right)^{5/2}}\right]=\nonumber\\
	&=-\frac{1}{2}\int\diff\rr\left[ \left(\frac{\delta \Psi}{\delta \rho^*}, \frac{\delta \Psi}{\delta e^*}\right) \cdot \right.\nonumber\\
	&\quad\left.
	e^{-m \rho^*/k_B} \left(\frac{2 \pi m k_B}{h^2 e^*}\right)^{3/2}
	\begin{pmatrix}
		-\frac{m^2}{k_B} & -\frac{3}{2}\frac{m}{e^*}\\
		-\frac{3}{2}\frac{m}{e^*} & -\frac{15}{4}\frac{k_B}{(e^*)^2}
	\end{pmatrix}
	\cdot
	\begin{pmatrix}
		\frac{\delta \Psi}{\delta \rho^*}\\
		\frac{\delta \Psi}{\delta e^*}
	\end{pmatrix}\right],
\end{align}
which is an equation for $\Psi$ (being equal to $\dXi$).
Let us now investigate properties of this equation. Firstly, the equation can be altered by adding an arbitrary term of form $\nabla\cdot \BB$, i.e. divergence of a vector field, without changing the content of the equation, since boundary terms are dropped in the functional derivatives since the beginning. This is a weak point of this approach and the only remedy is specifying boundary conditions at the beginning and keeping the boundary terms. Also the left hand side contains second spatial derivatives, but not in the form of Laplacians, but rather Sobolev-like norms. 

Further simplification of the Hamilton-Jacobi equation by taking $e^*=\mathrm{const.}=T^{-1}$ and $\Psi_{e^*}$ leads to 
\begin{equation}
	\int\diff\rr \frac{k_B T}{m} (\nabla\rho^*)^2 = \int\diff \rr\left(\frac{\delta \Psi}{\delta \rho^*}\right)^2 = \Big|\Big| \frac{\delta \Psi}{\delta \rho^*}\Big|\Big|^2,
\end{equation}
which means that we get an equation for the $L^2$-norm of the derivative of $\Psi$. Boundary conditions should be specified in order to gain more information from the Hamilton-Jacobi equation.

\section{Conclusion}
Consider GENERIC time evolution on a manifold $\MMM$ parameterized by state variables $\xx\in\MMM$. Assuming a linear projection $\dpi$ from $\MMM$ to a lower-dimensional manifold $\NNN$ parametrized by $\yy\in\NNN$, exact evolution on $\NNN$ can be obtained by projection of evolution on $\MMM$, $\dot{\yy} = \dpi(\dot{\xx})$, i.e. at each time-step state variables $\xx$ are updated by their exact evolution and then projected to $\NNN$.
But if we seek
evolution on $\NNN$ expressed only in terms of $\yy$, how to find it?

A possible answer to this question is provided by the lack-of-fit reduction method \cite{Turkington}, where Liouville equation for probability densities is reduced to less detailed evolution for moments of the distribution function in the GENERIC form. In the current work the reduction is generalized so that evolution on the upper (more detailed) level of description can be in GENERIC form generated by arbitrary entropy, arbitrary Poisson bracket, arbitrary energy and arbitrary dissipation potential. The original Liouville equation is then a particular choice of the Liouville Poisson bracket, Boltzmann entropy and no dissipation potential. The resulting evolution on the less detailed level is then again in the GENERIC form.
This generalization is illustrated on the reduction of a reversible harmonic oscillator to dissipative evolution of its position, and on the reduction from the Vlasov equation to hydrodynamics and mechanical equilibrium of ideal gases.

\section*{Acknowledgment}
This research has been supported partially by the Natural Sciences and Engineering Research Council of Canada, Grants 3100319 and 3100735.
This work was also supported by Czech Science Foundation, project no.  17-15498Y, and by Charles University Research program No. UNCE/SCI/023.

We are grateful to Karel T\r{u}ma for discussing convexity of dissipation potentials and to Alexander Mielke for discussing the invariance of gradient dynamics on change of variables.


\appendix

\section{Details of the static reduction}\label{sec.static.details}
This appendix contains important relations between state and conjugate variables and entropies on both the upper and lower levels.

\subsection{Transformation $\xx\leftrightarrow \xx^*$}
Variables $\xx$ and $\xx^*$ are related by Legendre transform
\begin{equation}
 \frac{\partial}{\partial x^i}\left(-\uS + x^i x^*_i \right) = 0
\end{equation}
where $\uS(\xx)$ is entropy on the higher level. By solving this last equation, one obtains
\begin{equation}\label{eq.uS.d}
 x^*_i = \frac{\partial \uS}{\partial x^i} \mbox{ and the dependence } \xx(\xx*).
\end{equation}
By differentiating it follows that
\begin{equation}\label{eq.uS.dd}
 \frac{\partial^2 \uS}{\partial x^i \partial x^j}= \frac{\partial x^*_i}{\partial x^j}.
\end{equation}
The conjugate entropy on the higher level is given by
\begin{equation}\label{eq.uS*}
 \uS^* = -\uS(\xx(\xx^*)) + x^i(\xx^*) x^*_i.
\end{equation}

The inverse transformation is carried out by Legendre transform
\begin{equation}
 \frac{\partial}{\partial x^*_i}\left(-\uS^* + x^*_i x^i \right) = 0
\end{equation}
where $\uS^*(\xx)$ is the conjugate entropy on the higher level, given by \eqref{eq.uS*}. By solving this last equation, one obtains
\begin{equation}\label{eq.uS*.d}
 x^i = \frac{\partial \uS^*}{\partial x^*_i}
\end{equation}
and by differentiating that
\begin{equation}
 \frac{\partial ^2 \uS^*}{\partial x^*_i \partial x^*_j}= \frac{\partial x^i}{\partial x^*_j}.
\end{equation}
By comparing with \eqref{eq.uS.dd}, it follows that
\begin{equation}\label{eq.uS.uS*.dd}
 \frac{\partial^2 \uS}{\partial x^i \partial x^j} = \left(\frac{\partial^2 \uS^*}{\partial \xx^* \partial \xx^*}\right)^{-1}_{ij},
\end{equation}
which means that the second differentials of $d^2 \uS$ and $d^2 \uS^*$ are inverse to each other.

Let $\uE(\xx)$ be energy on the upper level of description.
Conjugate energy on the higher level is introduced as
\begin{equation}
 \uE^* = \uE(\xx(\xx^*)),
\end{equation}
which means that
\begin{equation}
 \frac{\partial \uE}{\partial x^i} = \frac{\partial x^*_j}{\partial x^i}\frac{\partial \uE^*}{\partial x^*_j}.
\end{equation}
Note that this is \emph{not} a Legendre transformation.

\subsection{Transformation $\yy\leftrightarrow \yy^*$}\label{sec.yy.yy*}
Variables $\yy^*$ and $\yy$ are related by Legendre transform
\begin{equation}
 \frac{\partial}{\partial \yy^*}\left(-\dS^* + y^*_a y^a \right) = 0
\end{equation}
where $\dS^*(\yy)$ is the conjugate entropy on the lower level, given by \eqref{eq.dS*}. By solving this last equation, one obtains
\begin{equation}\label{eq.y*.y}
 y^a = \frac{\partial \dS^*}{\partial y^*_a} \mbox{ and the dependence } \yy^*(\yy).
\end{equation}
Differentiating with respect to $\yy^*$ leads to
\begin{equation}\label{eq.dS*.dd}
 \frac{\partial^2 \dS^*}{\partial y^*_a \partial y^*_b} = \frac{\partial y^a}{\partial y^*_b}.
\end{equation}
Entropy on the lower level is then simply the Legendre transform
\begin{equation}
 \dS = -\dS^*(\yy^*(\yy)) + y^*_a(\yy) y^a.
\end{equation}

The inverse transformation is carried out by Legendre transform
\begin{equation}
 \frac{\partial}{\partial y^i}\left(\dS + y^i y^*_i \right) = 0.
\end{equation}
By solving this last equation, one obtains
\begin{equation}\label{eq.dS.d}
 y^*_a = \frac{\partial \dS}{\partial y^a},
\end{equation}
and by differentiating that
\begin{equation}\label{eq.dS.dd}
 \frac{\partial^2 \dS}{\partial y^a \partial y^b}= \frac{\partial y^*_a}{\partial y^b}.
\end{equation}
By comparing with \eqref{eq.dS*.dd}, it follows that
\begin{equation}\label{eq.dS.dS*.dd}
 \frac{\partial^2 \dS}{\partial y^a \partial y^b} = \left(\frac{\partial^2 \dS^*}{\partial \yy^* \partial \yy^*}\right)^{-1}_{ab}.
\end{equation}

Let us also derive an identity useful later.
\begin{equation}\label{eq.yy*.y*y}
 \frac{\partial y^c}{\partial y^*_a} = \frac{\partial^2 \dS^*}{\partial y^*_a \partial y^*_c} = \left(\left(\frac{\partial^2 \dS}{\partial \yy \partial \yy}\right)^{-1}\right)^{ac} = \left(\left(\frac{\partial \yy^*}{\partial \yy}\right)^{-1}\right)^{ac} = \frac{\partial y^a}{\partial y^*_c}.
\end{equation}

\subsection{MaxEnt embedding $\yy^* \rightarrow \xx$}\label{sec.MaxEnt}
Mapping from conjugate variables on the lower level, $\yy^*$, to state variables on the higher level, $\xx$, is given by maximization of entropy $\uS$ with constraints given by the lower conjugate variables, i.e. by a generalized Legendre transform
\begin{equation}
 \frac{\partial}{\partial \xx} \left(-\uS + y^*_a \dpi^a(\xx)\right) = 0,
\end{equation}
which gives
\begin{equation}\label{eq.uS.d.y}
 \frac{\partial \uS}{\partial x^i}(\yy^*) = y^*_a \frac{\partial \dpi^a}{\partial x^i}\Big|_{\xx(\yy^*)}\mbox{ and the dependence }\xx(\yy^*).
\end{equation}
Note that if $\dpi$ is one-to-one mapping, the generalized Legendre transformation can be inverted and can be seen as a change of variables. On the other hand, if $\dpi$ is not one-to-one, e.g. a projection, then the generalized Legendre transformation can not be inverted.

The conjugate lower entropy is
\begin{equation}\label{eq.dS*}
 \dS^* = -\uS(\xx(\yy^*)) + y^*_a \dpi^a(\xx(\yy^*)).
\end{equation}
From Eq. \eqref{eq.uS.d.y} it follows by differentiation with respect to $y^b$ that
\begin{equation}
	\frac{\partial^2 \uS}{\partial x^i \partial x^j}\Big|_{\xx(\yy^*(\yy))} \frac{\partial x^j}{\partial y^*_a}\frac{\partial y^*_a}{\partial y^b}
	= \frac{\partial y^*_a}{\partial y^b} \frac{\partial y^a}{\partial x^i}\Big|_{\xx(\yy^*(\yy))},
\end{equation}
where the linearity of $\dpi$ was used.

Note that we can now apply Legendre transformation to the lower conjugate entropy $\dS^*$, giving the dependence $\yy^*(\yy)$. This dependence can be then plugged into the relation $\xx(\yy^*)$, giving the MaxEnt embedding $\xx(\yy^*(\yy))$ of $\NNN$ into $\MMM$. MaxEnt can be thus seen as a sequence of a generalized Legendre transformation and a standard Legendre transformation.

It is also useful to keep in mind that
\footnote{Note, on the other hand, that the commutation diagram $\xx(\yy^*(\dpi(\xx)))\neq \xx$ does not work, which means that the MaxEnt embedding of $\NNN$ in $\MMM$ does not always reconstruct the original position in $\MMM$.}
\begin{equation}
\dpi \circ \xx(\yy^*(\yy)) = \yy \quad\forall\yy.
\end{equation}
Differentiation of this relation then gives
\begin{equation}
\frac{\partial \dpi^a}{\partial x^i}\Big|_{\xx(\yy^*)} \frac{\partial x^i}{\partial y^*_c}\frac{\partial y^*_c}{\partial y^b} = \delta^a_b,
\end{equation}
which after multiplication by $\frac{\partial y^b}{\partial y^*_d}$ gives
\begin{equation}\label{eq.pi.diff}
	\frac{\partial \dpi^a}{\partial x^i}\Big|_{\xx(\yy^*)} \frac{\partial x^i}{\partial y^*_d} = \frac{\partial y^a}{\partial y^*_d}.
\end{equation}

Finally, we shall derive a relation between second differentials of the lower conjugate entropy $\dS^*$ and the higher-level entropy $\uS$. Taking derivative of definition \eqref{eq.dS*}, we obtain
\begin{equation}
	\frac{\partial \dS^*}{\partial y^*_b} \stackrel{\mbox{Eq.\eqref{eq.pi.diff}}}{=}
	- \frac{\partial \uS}{\partial x^i}\Big|_{\xx(\yy^*)} \frac{\partial x^i}{\partial y^*_b}
	+ \dpi^b(\xx(\yy^*))
	+ y^*_a \frac{\partial y^a}{\partial y^*_b}    
\end{equation}
Taking another derivative with respect to $y^*_c$ leads to
\begin{eqnarray}
	\frac{\partial^2 \dS^*}{\partial y^*_c \partial y^*_b} &=&
	-\frac{\partial^2 \uS}{\partial x^j\partial x^i}\Big|_{\xx(\yy^*)}\frac{\partial x^j}{\partial y^*_c}\frac{\partial x^i}{\partial y^*_b}
	-\frac{\partial \uS}{\partial x^i}\Big|_{\xx(\yy^*)} \frac{\partial^2 x^i}{\partial y^*_c\partial y^*_b} \nonumber\\
	&&+\frac{\partial \dpi^b}{\partial x^i}\Big|_{\xx(\yy^*)} \frac{\partial x^i}{\partial y^*_c}
	+\frac{\partial y^c}{\partial y^*_b}
	+y^*_a \frac{\partial^2 y^a}{\partial y^*_c \partial y^*_b}\nonumber\\
	&\stackrel{\mbox{Eq. \eqref{eq.uS.d.y}}}{=}&
	-\frac{\partial^2 \uS}{\partial x^j\partial x^i}\Big|_{\xx(\yy^*)}\frac{\partial x^j}{\partial y^*_c}\frac{\partial x^i}{\partial y^*_b}
	-\frac{\partial \uS}{\partial x^i}\Big|_{\xx(\yy^*)} \frac{\partial^2 x^i}{\partial y^*_c\partial y^*_b} \nonumber\\
	&&+y^*_a\frac{\partial^2 y^a}{\partial y^*_b \partial y^*_c}
	+2 \frac{\partial^2 \dS^*}{\partial y^*_b \partial y^*_c},
\end{eqnarray}
from which it follows that
\begin{eqnarray}
	\frac{\partial^2 \uS}{\partial x^j\partial x^i}\Big|_{\xx(\yy^*)}\frac{\partial x^j}{\partial y^*_c}\frac{\partial x^i}{\partial y^*_b}  &=&
	\frac{\partial^2 \dS^*}{\partial y^*_c \partial y^*_b}
	-y^*_a \frac{\partial \dpi^a}{\partial x^i}\Big|_{\xx(\yy^*)} \frac{\partial^2 x^i}{\partial y^*_b \partial y^*_c}
	+ y^*_a \frac{\partial^2 y^a}{\partial y^*_b\partial y^*_c}\nonumber\\
	&=&
	\frac{\partial^2 \dS^*}{\partial y^*_c \partial y^*_b}
	-y^*_a \left(\frac{\partial}{\partial y^*_b}\left(\frac{\partial \dpi^a}{\partial x^i}\Big|_{\xx(\yy^*)} \frac{\partial x^i}{\partial y^*_c}\right)
	- \frac{\partial^2 y^a}{\partial y^*_b\partial y^*_c}\right)\nonumber\\
	&\stackrel{\mbox{Eq. \eqref{eq.pi.diff}}}{=}&
	\frac{\partial^2 \dS^*}{\partial y^*_c \partial y^*_b}.\label{eq.d2uS.d2dS*}
\end{eqnarray}
This identity is useful in the variation of the lack-of-fit Lagrangian.

\subsection{Projection $\xx\rightarrow \yy$} \label{sect.Projection}
Passage from the higher level to the lower level is given by projection \eqref{eq.projection}. The lower level entropy $\dS$ is given by Legendre transformation of $\dS^*$ in Sec. \ref{sec.yy.yy*}

Let $\dE$ be energy on the lower lever implied by the upper-level energy, i.e.
\begin{equation}
	\dE(\yy) = \uE(\xx(\yy^*(\yy)))\quad \forall \yy
\end{equation}
Conjugate lower-level energy is then
\begin{equation}
 \dE^*(\yy^*) = \dE(\yy(\yy^*)),
\end{equation}
and it holds that
\begin{equation}\label{eq.dE.d}
	\frac{\partial \dE}{\partial y^a} = \frac{\partial \uE}{\partial x^i}\Big|_{\xx(\yy*(\yy))} \frac{\partial x^i}{\partial y^*_b}\Big|_{\yy^*(\yy)} \frac{\partial y^*_b}{\partial y^a},
\end{equation}
which shows how to calculate gradient of the lower energy.

\subsection{Conjugate embedding $\yy^* \rightarrow \xx^*$}
Variables $\yy$ are obtained from variables $\xx$ by projection $\dpi$. The dual mapping then gives the dependence $\xx^*(\yy^*)$ by requiring the following compatibility
\begin{equation}
 \xx(\yy^*) = \xx(\xx^*(\yy^*)).
\end{equation}

Let us now prove a useful identity for the second differential of the lower conjugate entropy.
Using
\begin{eqnarray}
	\frac{\partial x^i}{\partial y^*_a}\frac{\partial}{\partial y^*_b} \frac{\partial \uS}{\partial x^i}
	= \frac{\partial x^i}{\partial y^*_a} \frac{\partial^2 \uS}{\partial x^i \partial x^j}\Big|_{\xx(\yy^*)} \frac{\partial x^j}{\partial y^*_b}\nonumber\\
\end{eqnarray}
and relation \eqref{eq.d2uS.d2dS*}, we obtain that
\begin{equation}\label{eq.A}
\frac{\partial x^i}{\partial y^*_a}\frac{\partial x^*_i}{\partial y^*_b} = \frac{\partial^2 \dS^*}{\partial y^*_a \partial y^*_b},
\end{equation}
which is used in the calculation of variation of action $\Psi$.

\section{Convexity of dissipation potential}\label{sec.convex}
Dissipation potentials are usually considered as convex functionals of the conjugate variables. This restriction is not necessary, since the second law is implied already by monotonicity of derivative of the dissipation potential (see \cite{nonconvex}), but convexity implies monotonicity, as we show in the following text.

\subsection{Convexity implies monotonicity}\label{sec.mon}
Consider a convex functional $\Xi$ on a function space elements of which are denoted by $X$, $Y$, etc. Convexity means that
\begin{equation}\label{eq.Xi.convex}
	\Xi(\alpha X+(1-\alpha) Y) \leq \alpha \Xi(X) + (1-\alpha) \Xi(Y) \qquad\forall \alpha\in[0,1].
\end{equation}
Our goal is to prove inequality
\begin{equation}\label{eq.Xi.mono}
	(\Xi'|_X - \Xi'|_Y)\cdot(X-Y)\geq 0,
\end{equation}
where $\Xi'$ stands for the Fréchet derivative of $\Xi$. When $X$ and $Y$ are real numbers and $\Xi$ is a real-valued function, the inequality can be easily seen graphically. We will however consider the general case when $X$ and $Y$ are fields and $\Xi$ is a smooth functional of them.

Define first a function from real numbers to real numbers
\begin{equation}
	\psi(\alpha) = \Xi(\alpha X +(1-\alpha)Y) - \alpha \Xi(X) - (1-\alpha)\Xi(Y).
\end{equation}
This function is negative for all $\alpha\in[0,1]$ due to convexity of $\Xi$ and is equal to zero at $\alpha=0$ and $\alpha = 1$. Therefore, there must be a point $\alpha_0 \in (0,1)$ such that
\begin{equation}
	\frac{d \psi}{d \alpha}\Big|_{\alpha_0} = 0.
\end{equation}
We thus know behavior of the function $\psi$ at three points.

Is the derivative of $\psi$ always negative for $\alpha<\alpha_0$ and positive for $\alpha>\alpha_0$? The answer is affirmative, but it requires a little work to prove. If we show that $\psi$ is a convex function, then the inequalities
\begin{equation}\label{eq.psi.ineq}
	\frac{d \psi}{d \alpha}<0 \quad\forall\alpha<\alpha_0
	\qquad\mbox{and}\qquad
	\frac{d \psi}{d \alpha}>0 \quad\forall\alpha>\alpha_0
\end{equation}
follow easily. Convexity of $\psi$ is the inequality
\begin{eqnarray}
	\psi(\xi\alpha + (1-\xi)\beta) &=&
	\Xi((\xi\alpha + (1-\xi)\beta)X + (1-(\xi\alpha + (1-\xi)\beta))Y) \nonumber\\
	&&-(\xi\alpha + (1-\xi)\beta)\Xi(X)-(1-(\xi\alpha + (1-\xi)\beta))\Xi(Y)\nonumber\\
	&\leq& \xi \Xi(\alpha X+(1-\alpha)Y) + (1-\xi)\Xi(\beta X+(1-\beta)Y)\nonumber\\
	&&-(\xi\alpha + (1-\xi)\beta)\Xi(X)-(1-(\xi\alpha + (1-\xi)\beta))\Xi(Y)\nonumber\\
	&=&\xi \psi(\alpha)+(1-\xi)\psi(\beta).
\end{eqnarray}
Therefore, inequalities \eqref{eq.psi.ineq} are valid, and can be rewritten as
\begin{subequations}
\begin{eqnarray}
	\Xi'(\alpha X+(1-\alpha)Y) \cdot(X-Y)-\Xi(X)+\Xi(Y) \leq 0 \qquad\forall \alpha<\alpha_0\\
	\Xi'(\alpha X+(1-\alpha)Y) \cdot(X-Y)-\Xi(X)+\Xi(Y) \geq 0 \qquad\forall \alpha>\alpha_0.
\end{eqnarray}
\end{subequations}
In particular, for $\alpha=0$ and $\alpha = 1$ we get
\begin{subequations}
\begin{eqnarray}
	\Xi'|_Y \cdot(X-Y) \leq \Xi(X)-\Xi(Y)\\
	\Xi'|_X \cdot(X-Y) \geq \Xi(X)-\Xi(Y).
\end{eqnarray}
\end{subequations}
By adding these two inequalities we obtain that
\begin{equation}
	(\Xi'(X)-\Xi'(Y))\cdot(X-Y) \geq \Xi(X)-\Xi(Y)-\Xi(X)+\Xi(Y)=0,
\end{equation}
which is the desired inequality telling that Fréchet derivative $\Xi'$ is a monotone operator, see \cite{Roubicek}, Sec. 4.1., for a rigorous derivation.

In particular, if choose $Y=0$, we obtain the second law of thermodynamics,
\begin{equation}
	\dot{S} = X \cdot \Xi'(X) \geq 0
\end{equation}
where $X$ is the thermodynamic force, e.g. conjugate variable $X = S_x$.
This can summarized in the following Lemma
\begin{lemma}
	If $\Xi(X)$ is a convex functional, then its derivative is a monotone operator and, in particular, gradient dynamics satisfies the second law of thermodynamics.
\end{lemma}

\subsection{Convexity of the dissipation potential for heat conduction}\label{app.Xi.heat}
The aim is to show that dissipation potential generating Fourier heat conduction,
\begin{equation}\label{eq.Xi.heat}
	\Xi(e^*) = \int_{\Omega} \frac{1}{2}\lambda \nabla e^* \cdot \nabla e^*,
\end{equation}
is convex, which means that
\begin{eqnarray}
	\Xi(\alpha e^* + (1-\alpha) f^*)&=& \int_\Omega \frac{1}{2}\lambda \left(\alpha \nabla e^*+(1-\alpha)\nabla f^*\right)^2\nonumber\\
	&=& \int_\Omega \lambda \alpha^2\frac{1}{2} (\nabla e^*)^2
	 +\int_\Omega \lambda (1-\alpha)^2\frac{1}{2} (\nabla f^*)^2 \nonumber\\
	 &&+\alpha(1-\alpha)\int_\Omega \nabla e^*\cdot \nabla f^*\nonumber\\
	&\leq& \int_\Omega \lambda \alpha^2\frac{1}{2} (\nabla e^*)^2
	 +\int_\Omega \lambda (1-\alpha)^2\frac{1}{2} (\nabla f^*)^2 \nonumber\\
	 &&+\alpha(1-\alpha)\sqrt{\int_\Omega \lambda(\nabla e^*)^2} \cdot \sqrt{\int_\Omega \lambda (\nabla f^*)^2}\nonumber\\
	&\leq& \int_\Omega \lambda \alpha^2\frac{1}{2} (\nabla e^*)^2
	 +\int_\Omega \lambda (1-\alpha)^2\frac{1}{2} (\nabla f^*)^2 \nonumber\\
	 &&+\alpha(1-\alpha)\left(\int_\Omega \frac{1}{2}\lambda(\nabla e^*)^2+\int_\Omega \frac{1}{2}\lambda (\nabla f^*)^2\right)\nonumber\\
	 &=& \alpha \Xi(e^*) + (1-\alpha) \Xi(f^*).
\end{eqnarray}
The first inequality is an application of H{\" o}lder's inequality while the second follows from Young's inequality. The dissipation potential is convex.

\section{Coordinate invariance of gradient dynamics}\label{sec.GD}
In this section we show that gradient dynamics is of course invariant with respect to bijective change of variables, as shown in \cite{mielke-generic} in the case of quadratic dissipation potentials.
Consider a set of state variables $\xx$ and a new set of state variables $\nxx(\xx)$, and assume that the relation between them is bijective so that we also have $\xx(\nxx)$.

\subsection{Transformation of conjugate variables}
Having entropy as a functional of $\xx$, $S(\xx)$, entropy in the new variables is $\nS(\nxx)=S(\xx)$. Taking derivative of this relation leads to
\begin{equation}
	\frac{\partial S}{\partial x^i} = \frac{\partial \nS}{\partial \nx^j}\frac{\partial \nx^j}{\partial x^i}
	\qquad\mbox{and}\qquad
	\frac{\partial \nS}{\partial \nx^i} = \frac{\partial S}{\partial x^j}\frac{\partial x^j}{\partial \nx^i},
\end{equation}
or
\begin{equation}\label{eq.xx*.nxx*}
	x^*_i = \frac{\partial \nx^j}{\partial x^i}\nx^*_j
	\qquad\mbox{and}\qquad
	\nx^*_i = \frac{\partial x^j}{\partial \nx^i} x^*_j.
\end{equation}
Derivatives of these relation with respect to the conjugate variables are
\begin{equation}\label{eq.dx*.dnx*}
	\frac{\partial x^*_i}{\partial \nx^*_k} = \frac{\partial \nx^k}{\partial x^i}
	\qquad\mbox{and}\qquad
	\frac{\partial \nx^*_i}{\partial x^*_k} = \frac{\partial x^k}{\partial \nx^i},
\end{equation}
which represent the transformation rules for the conjugate variables.

\subsection{Transformation of dissipation potential}
Let us now assume gradient dynamics in the $\xx$ variables,
\begin{equation}\label{eq.q.grad}
	\dot{x}^i = \frac{\partial \Xi}{\partial x^*_i}\Big|_{\xx^* = S_\xx}.
\end{equation}
The new dissipation potential (in the new variables) is defined by
\begin{equation}\label{eq.nXi}
	\nXi(\nxx^*) = \Xi(\xx^*(\nxx^*)).
\end{equation}
Evolution of the new variables implied by evolution of the old ones, Eq. \eqref{eq.q.grad}, is
\begin{eqnarray}
	\dot{\nx}^i &=& \frac{\partial \nx^i}{\partial x^j}\frac{\partial \Xi}{\partial x^*_j}
	= \frac{\partial \nx^i}{\partial x^j} \frac{\partial \nXi}{\partial \nx^*_k} \frac{\partial \nx^*_k}{\partial x^*_j}\nonumber\\
	&=&
	\frac{\partial \nx^i}{\partial x^j} \frac{\partial \nXi}{\partial \nx^*_k} \frac{\partial x^j}{\partial \nx^k}
	= \frac{\partial \nXi}{\partial \nx^*_i},
\end{eqnarray}
which is again gradient dynamics in the new variables $\nxx$. Thus we have proved the following lemma:
\begin{lemma}\label{lemma.grad.inv}
	Gradient dynamics is invariant with respect to change of coordinates, and conjugate variables transform according to relations \eqref{eq.xx*.nxx*}.
\end{lemma}

\subsection{Special case: quadratic dissipation potential}
Assume now that the dissipation potential is quadratic,
\begin{equation}\label{eq.Xi.M}
	\Xi(\xx^*) = \frac{1}{2} x^*_i M^{ij} x^*_j.
\end{equation}
The symmetric positive definite operator $\MM$ is called dissipative matrix. The new dissipation potential \eqref{eq.nXi} is, using transformation rules \eqref{eq.xx*.nxx*},
\begin{equation}\label{eq.nXi.M}
	\nXi(\nxx^*) = \frac{1}{2} \nx^*_k \underbrace{\frac{\partial \nx^k}{\partial x^i} M^{ij} \frac{\partial \nx^l}{\partial x^j}}_{\stackrel{def}{=}\nM^{kl}} \nx^*_l,
\end{equation}
where the new dissipative matrix was defined.

Gradient dynamics \eqref{eq.q.grad} becomes in the quadratic case
\begin{equation}
	\dot{x}^i = M^{ij} x^*_j,
\end{equation}
and the implied evolution of the new variables reads
\begin{equation}
	\dot{\nx}^i = \nM^{ij} \nx^*_j,
\end{equation}
which has of course the same structure as the evolution of $\xx$, being a special case of Lemma \eqref{lemma.grad.inv}.

\subsection{Invariance of convexity}
Assume that $\Xi$ is a convex functional of $\xx^*$, Eq. \eqref{eq.Xi.convex}. Then from the definition \eqref{eq.nXi} it follows that
\begin{equation}
	\nXi(\nxx^*(\alpha \xx^*_1+(1-\alpha) \xx^*_2)) \leq \alpha \nXi(\nxx^*(\xx^*_1)) + (1-\alpha) \nXi(\nxx^*(\xx^*_2)) \qquad\forall \alpha\in[0,1].
\end{equation}
From the linearity of transformation relations \eqref{eq.xx*.nxx*} it follows that
\begin{equation}
	\nXi(\alpha \nxx^*(\xx^*_1)+(1-\alpha) \nxx^*(\xx^*_2)) \leq \alpha \nXi(\nxx^*(\xx^*_1)) + (1-\alpha) \nXi(\nxx^*(\xx^*_2)) \qquad\forall \alpha\in[0,1],
\end{equation}
which (by the regularity of mapping $\nxx(\xx)$) yields
\begin{equation}
	\nXi(\alpha \nxx^*_1+(1-\alpha) \nxx^*_2) \leq \alpha \nXi(\nxx^*_1) + (1-\alpha) \nXi(\nxx^*_2) \qquad\forall \alpha\in[0,1],
\end{equation}
which is convexity of $\nXi$.

\begin{lemma}
	Dissipation potential $\Xi$ is a convex functional of $\xx^*$ if and only if the transformed dissipation potential $\nXi$ is a convex functional of $\nxx^*$.
\end{lemma}

In Sec. \ref{sec.convex} it was shown that convexity implies that the respective gradient dynamics fulfills the second law of thermodynamics (entropy grows). Is this property invariant with respect to the change of coordinates? The answer is affirmative as
\begin{equation}
	x^*_i \frac{\partial \Xi}{\partial x^*_i}
	= \frac{\partial \nx^j}{\partial x^i} x^*_j \frac{\partial \nXi}{\partial \nx^*_k} \frac{\partial \nx^*_k}{\partial x^*_i}
	\stackrel{\mbox{Eq. \eqref{eq.dx*.dnx*}}}{=} \nx^*_j \frac{\partial \nXi}{\partial \nx^*_j}.
\end{equation}
\begin{lemma}
	Dissipation potential $\Xi$ satisfies the second law of thermodynamics if and only if the transformed dissipation potential $\nXi$ does.
\end{lemma}

\subsection{Entropic vs. energetic representation}
Let us now demonstrate the invariance of gradient dynamics with respect to transformations of state variables on non-isothermal particle mechanics, where the particle is equipped energy, i.e. with state variables $\nxx=(q,p,e)$. This is the entropic representation, where the fundamental thermodynamic relation, see \cite{callen}, is given as the function of entropy on the state variables $\nxx$, e.g.
\begin{subequations}\label{eq.se}
\begin{equation}
	s(q,p,e) = \sigma\left(e - \frac{p^2}{2m} - V(q)\right).
\end{equation}
The function $\sigma(\bullet)$ is the inverse of function $\epsilon(s)$, i.e. $\sigma(\epsilon(s)) = s$, expressing the contribution of internal energy to the total energy,
\begin{equation}
	e = \frac{p^2}{2m} + V(q) + \epsilon(s).
\end{equation}
\end{subequations}
But let us now proceed in full generality, i.e. regardless the concrete expressions \eqref{eq.se}.

Instead of the entropic representation, it is often useful to formulate the equations in the energetic representation, where entropy is among the state variables. This is particularly useful when writing down the reversible Hamiltonian evolution. State variables in the energetic representation are $\xx=(q,p,s)$. Since the transformation between $\nxx$ and $\xx$ is obvious, let us focus on the transformation between the conjugate variables, $\nxx^*$ and $\xx^*$. From relations \eqref{eq.xx*.nxx*} we obtain
\begin{subequations}
	\begin{eqnarray}
		q^* &=& \frac{\partial \nx^i}{\partial q} \nx^*_i
		= \nq^* + \left(\frac{\partial e}{\partial q}\right)_{p,s} e^*\\
		p^* &=& \frac{\partial \nx^i}{\partial p} \nx^*_i
		= \np^* + \left(\frac{\partial e}{\partial p}\right)_{q,s} e^*,\\
		s^* &=& \frac{\partial \nx^i}{\partial s} \nx^*_i
		=\left(\frac{\partial e}{\partial s}\right)_{q,p} e^*,
	\end{eqnarray}
\end{subequations}
and the inverse
\begin{subequations}
	\begin{eqnarray}
		\nq^* &=& \frac{\partial x^i}{\partial q}x^*_i
		= q^* + \left(\frac{\partial s}{\partial q}\right)_{p,e} s^*\\
		\np^* &=& \frac{\partial x^i}{\partial p}x^*_i
		= p^* + \left(\frac{\partial s}{\partial p}\right)_{q,e} s^*,\\
		e^* &=& \frac{\partial x^i}{\partial e}x^*_i
		=\left(\frac{\partial s}{\partial e}\right)_{q,p} s^*.
	\end{eqnarray}
\end{subequations}
These two sets of transformations are compatible due to the implicit function theorem,
\begin{equation}
	\left(\frac{\partial s}{\partial q}\right)_{p,e} =
	 -\frac{\left(\frac{\partial e}{\partial q}\right)_{p,s}}{\left(\frac{\partial e}{\partial s}\right)_{q,p}}
	 \quad\mbox{and}\quad
	\left(\frac{\partial s}{\partial p}\right)_{q,e} =
	 -\frac{\left(\frac{\partial e}{\partial p}\right)_{q,s}}{\left(\frac{\partial e}{\partial s}\right)_{q,p}}.
\end{equation}
These transformations can be considered as generally applicable also for other sets of state variables, where the relations are analogical.

Now using the particular relations \eqref{eq.se}, which allow for explicit identification of conjugate variables with derivatives of entropy, the conjugate variables become
\begin{subequations}
	\begin{equation}
		\nq^* = -\frac{\partial \sigma}{\partial e} \frac{\partial V}{\partial q},\quad
		\np^* = -\frac{\partial \sigma}{\partial e} \frac{p}{m},\quad\mbox{and}\quad
		e^* = \frac{\partial \sigma}{\partial e}
	\end{equation}
	in the entropic representation,
\end{subequations}
and
\begin{subequations}
	\begin{equation}
		q^* = 0,\quad
		p^* = 0,\quad\mbox{and}\quad
		s^* = 1
	\end{equation}
	in the energetic representation.
\end{subequations}
We have thus found the general relations for conjugate variables in the energetic and entropic representations, as well as the particular expressions when a concrete energy (or entropy) is chosen.

Let us now turn to the evolution given by gradient dynamics. Consider quadratic gradient dynamics in the entropic representation \eqref{eq.nXi.M}, for instance with
\begin{equation}
	\nMM =
	\begin{pmatrix}
		0 & 0 & 0\\
		0 & \zeta & 0\\
		0 & 0 & 0
	\end{pmatrix}.
\end{equation}
The resulting equations are thus
\begin{subequations}
	\begin{eqnarray}
		\dot{q} &=& 0\\
		\dot{p} &=& \zeta \np^*\\
		\dot{e} &=& 0,
	\end{eqnarray}
	where the particle experiences linear friction and transforms kinetic energy to internal so that the total energy is conserved.
\end{subequations}
Note that this $\nMM$ matrix has no coupling among the state variables, only the momentum dissipates.

In the energetic representation, the dissipation potential becomes
\begin{equation}
	\Xi(\xx^*) = \frac{1}{2}\zeta \left(p^* +\left(\frac{\partial s}{\partial p}\right)_{q,e} s^*\right)^2,
\end{equation}
which corresponds to dissipative matrix
\begin{equation}
	\MM =
	\begin{pmatrix}
		0 & 0 & 0\\
		0 & \zeta & \zeta \left(\frac{\partial s}{\partial p}\right)_{q,e}\\
		0 & \zeta \left(\frac{\partial s}{\partial p}\right)_{q,e} & \zeta \left(\frac{\partial s}{\partial p}\right)^2_{q,e}
	\end{pmatrix},
\end{equation}
which contains coupling between $p$ and $s$ as well as evolution in $s$,
\begin{subequations}
	\begin{eqnarray}
		\dot{q} &=& 0\\
		\dot{p} &=& \zeta p^* + \zeta \left(\frac{\partial s}{\partial p}\right)_{q,e}
		= \zeta \left(\frac{\partial s}{\partial p}\right)_{q,e}\\
		\dot{s} &=& \zeta \left(\frac{\partial s}{\partial p}\right)_{q,e} p^*
		+\zeta \left(\frac{\partial s}{\partial p}\right)^2_{q,e}
		=
		\zeta \left(\frac{\partial s}{\partial p}\right)^2_{q,e}
	\end{eqnarray}
	where the particle experiences linear friction and entropy is produced.
\end{subequations}

In summary, we have demonstrated the invariance of gradient dynamics with respect to the transformation between entropic and energetic transformation.

\subsection{Entropic and energetic representation in heat conduction}
Consider the field of energy density $e(\rr)$ and the dissipation potential \eqref{eq.Xi.heat} with coefficient $\lambda(T)$ dependent on temperature, $T = \left(\frac{\partial s}{\partial e}\right)^{-1}$. Gradient dynamics then reads
\begin{equation}
	\partial_t e = \frac{\delta \nXi}{\delta e^*(\rr)} = \frac{\delta}{\delta e^*(\rr)} \int_\Omega \dr \frac{1}{2} \lambda(T) (\nabla e^*)^2
	= -\nabla\cdot \left(\lambda(T) \nabla e^*\right),
\end{equation}
where $T^{-1}$ is to be substituted for $e^*$ in order to obtain equations in a closed form. This is the usual form of Fourier heat conduction,
\begin{equation}
	\partial_t e = - \nabla\cdot \qq, \qquad \qq = \lambda \nabla T^{-1}
\end{equation}
being the heat flux.

Let us now transform the dissipation potential to the energetic representation. Using Eqs. \eqref{eq.xx*.nxx*}, we obtain
\begin{equation}
	\Xi(s^*) = \int\dr \lambda(T) \left(\nabla \left(s^* T^{-1}\right)\right)^2.
\end{equation}
Evolution of the field of entropy density is then governed by
\begin{eqnarray}
	\partial_t s &=& \frac{\delta \Xi}{\delta s^*(\rr)} = -T^{-1}\nabla\cdot\left(\lambda(T) \nabla (s^* T^{-1})\right) \nonumber\\
	&=&-\nabla\cdot\left(T^{-1}\lambda\nabla T^{-1}\right) + \lambda(T) \left(\nabla T^{-1}\right)^2,
\end{eqnarray}
where the relation $s^* = 1$, which is always true, was used. Note that the first term on the right hand side represents divergence of $\qq/T$ while the second term is entropy production.

Gradient dynamics for heat conduction can be formulated in both entropic and energetic representations, and both formulations are compatible.

\end{document}